\documentclass[%
preprint,
amsmath,amssymb,
]{revtex4-1}

\usepackage[pdftex]{graphicx}
\usepackage{dcolumn}
\usepackage{bm}
\usepackage{braket}
\usepackage{color}
\usepackage{here}

\DeclareMathOperator*{\SumInt}{%
\mathchoice%
  {\ooalign{$\displaystyle\sum$\cr\hidewidth$\displaystyle\int$\hidewidth\cr}}
  {\ooalign{\raisebox{.14\height}{\scalebox{.7}{$\textstyle\sum$}}\cr\hidewidth$\textstyle\int$\hidewidth\cr}}
  {\ooalign{\raisebox{.2\height}{\scalebox{.6}{$\scriptstyle\sum$}}\cr$\scriptstyle\int$\cr}}
  {\ooalign{\raisebox{.2\height}{\scalebox{.6}{$\scriptstyle\sum$}}\cr$\scriptstyle\int$\cr}}
}

\def\m@thcombine#1#2{%
  \setbox0=\hbox{$#1$}
  \setbox1=\hbox{$#2$}
  \ifdim\wd0>\wd1
    \setbox0=\hbox to\wd1{\hss\box0\hss}
  \else
    \setbox1=\hbox to\wd0{\hss\box1\hss}
  \fi
  \mathop{\vcenter{
    \offinterlineskip\box0\box1}}}
\def\lesim{\m@thcombine<\sim}
\def\gesim{\m@thcombine>\sim}
\def\lessgtr{\m@thcombine<>}
\def\gtrless{\m@thcombine><}

\newcommand{\vecr}{\mbox{\boldmath $r$}}

\newcommand{\vecrs}{\mbox{\boldmath $r$}\sigma}

\newcommand{\del}{\partial}

\begin{document}
\title{
Continuum quasiparticle random-phase approximation for $(n,\gamma)$ reactions on neutron-rich nuclei: collectivity and
resonances in low-energy cross section
}
\author{Teruyuki Saito}
 \affiliation{Nuclear Data Center, Japan Atomic Energy Agency, Ibaraki 319-1195, Japan}
\author{Masayuki Matsuo}
\affiliation{Department of Physics, Faculty of Science. Niigata University, Niigata 950-2181, Japan}

\begin{abstract}
We formulate a microscopic theory to calculate cross section of the radiative neutron capture on neutron-rich nuclei using the continuum quasiparticle random-phase approximation. 
This formulation is designed to be applied to neutron-rich nuclei around the $r$-process path, for which the compound nuclear model may not be appropriate.
It takes into account effects of various excitation modes  such as the soft dipole excitation, the giant resonances, and the quasiparticle resonance in addition to the surface vibrations such as quadrupole and octupole modes.
We perform numerical calculations to demonstrate new features of the present theory, employing 
reactions $^{89}{\rm Ge}(n,\gamma)^{90}{\rm Ge}$ and $^{91}{\rm Zn}(n,\gamma)^{92}{\rm Zn}$  with $E1$ transitions populating the ground state, collective $2^{+}_{1}$ and $3^{-}_{1}$ states in $ ^{90}{\rm Ge}$ and $^{92}{\rm Zn}$.
With these examples, we discuss enhanced resonance contributions in the $(n,\gamma)$ reaction at low energy, which originate
from the quasiparticle resonance and the pygmy quadrupole resonance located  just above the one neutron separation energy, 
and from combination with the low-lying octupole vibrational state. 
\end{abstract}

\maketitle

\section{Introduction}

About half of the elements heavier than iron are believed to be synthesized in the $r$ process \cite{B2FH,Arnould2007,Cowan2021}.
The $r$-abundance estimation requires a large amount of nuclear data input such as nuclear masses, $(n,\gamma)$ and $(\gamma,n)$ reactions rates, $\beta$-decays rates, and the fission properties taking place near the neutron drip line far from the  $\beta$ stability. 
The cross sections and probabilities for these reactions and decays must be evaluated theoretically since the measurements for short-lived neutron-rich nuclei related with the $r$ process are quite difficult or impossible in most cases.
Recently the kilonova observation with a binary neutron star merger \cite{Abbott2017,Abbott2017_2} gave us the traces of the $r$-process nuclei synthesized \cite{Cowan2021,Pian2017,Kasen2017,Villar2017}.
With the progress of the $r$-process observation, reliable nuclear theories applicable to neutron-rich nuclei are demanded.

In the present study, we focus on the radiative neutron capture reaction.
It is usually described in terms of two different reaction mechanisms:
the compound nuclear (CN) process and the direct capture (DC) process.
For the CN process it  is assumed that a compound nucleus is formed through neutron capture, and the following $\gamma$ decay
occurs independently on the entrance channel of the capture.  
The cross section is usually evaluated by means of the Hauser and Feshbach statistical model \cite{Hauser1952}.
This CN mechanism is known to work well for stable nuclei with large neutron separation energy $S_{1n} \sim 8$ MeV,  
and is often applied to the $s$-process nucleosynthesis taking place on the stability line \cite{Kappeler2011}.
For the DC process, on the other hand, it is assumed that the $\gamma$ decay occurs directly from
the scattering state of the entrance channel to bound states of daughter nucleus.
Traditionally the DC is evaluated by potential models based on the independent particle picture.
In the case of the $r$ process occurring in neutron-rich nuclei, the separation energy is small $S_{1n} \sim 2$ MeV.
The statistical description of the CN process may not be appropriate because of the low level density at such small 
excitation energy \cite{Arnould2007,Cowan2021,Mathews1983}. Accordingly the DC process is taken into account  in the neutron
capture reaction for the $r$-process study \cite{Mathews1983, Rauscher1998, Bonneau2007, Chiba2008, Rauscher2010, Xu2012, Xu2014, Zhang2015, Sieja2021}.

Collective motions such as low-lying surface vibration modes and giant resonances are fundamental modes of excitation in nuclei.
In the case of neutron-rich nuclei far from the stability line, they additionally exhibit various exotic excitations not seen in stable nuclei such as  the pygmy dipole resonance or the soft dipole mode \cite{Tanihata1985, Hansen1987, Suzuki1990, Bertsch1991, Paar2007, Tanihata2013}, originating from the neutron halo, the neutron skin, or small neutron separation energy of last neutron(s). 
It has been argued that the pygmy dipole resonance influences the $r$-process nucleosynthesis \cite{Goriely1998}, and 
microscopically calculated $\gamma$-ray strength functions including effects of the pygmy dipole resonance
have been applied to the $r$-process studies \cite{Goriely2002, Goriely2004, Litvinova2009, Avdeenkov2011, Daoutidis2012, Xu2014, Tsoneva2015, Martini2016}. In the DC process models, however, effects of the collective excitations or resonances have been neglected.

The purpose of our study is to formulate a microscopic theory of the radiative neutron capture reaction which can be applicable to
the $r$ process and very neutron-rich nuclei, and to provide a better description of the cross section than the direct capture models by incorporating effects
of collective excitations and resonances originating from many-body correlations.

Previously we have formulated a prototype model which is based on the linear response theory, i.e., the continuum quasiparticle random-phase approximation (cQRPA)  through which the collectivity is taken into account \cite{Matsuo2015}. However the formulation describes only the $\gamma$ decay feeding to the 
ground state of the daughter nucleus. In another preceding paper \cite{Saito2023}, we developed a method to describe the $\gamma$ transition feeding the excited states, but it did not include pair correlation and hence  its can be applied only to closed shell nuclei.  
In the present study we implement the method of Ref.~\cite{Saito2023} to the prototype model \cite{Matsuo2015} so that the updated formulation
can be applied to open-shell neutron-rich nuclei.
As we show below, the new framework
enables us to describe resonances coupled to the entrance channel of the neutron capture as well as quadrupole and octupole surface vibrational states 
in the final states of the $\gamma$ decays.

We describe the radiative neutron capture as an inverse process of the $(\gamma,n)$ reaction,
and evaluate the cross section of the $(n,\gamma)$ reaction by means of the reciprocity theorem.
We employ the cQRPA approach, which can describe the photoabsorption reaction leading
to excited states including unbound single-particle configurations.
To utilize the reciprocity theorem, one needs to specify the initial and final states of the $(\gamma,n)$ reaction. This can be achieved by
applying the method of Zangwill and Soven \cite{Zangwill1980} to the cQRPA. 
Details of these formulations are presented in Sec.~\ref{theory}.  
We demonstrate new features of the present theory in Sec.~\ref{numerical_examples} and discuss two numerical examples of the $(n,\gamma)$ reaction on
neutron-rich odd-$N$ nuclei $^{91}{\rm Zn}$ and  $^{89}{\rm Ge}$ with $E1$ transitions, which populate the ground state as well as the low-lying quadrupole
$2^{+}_{1}$, and octupole $3^{-}_{1}$ collective states in $^{92}{\rm Zn}$ and $^{90}{\rm Ge}$.
These examples indicate possible presence of the pygmy quadrupole resonance and the quasiparticle resonance
that enhances the low-energy capture cross section. 
In Sec.~\ref{conclusions}, we draw conclusions.

\section{theory} \label{theory}

\subsection{initial and final configurations for $(n,\gamma)$ and $(\gamma, n)$ reactions}
In the present study, we describe a radiative neutron capture reaction on neutron-rich odd-$N$ nuclei with open-shell configuration.
We assume that the ground state of an odd-$N$ nucleus has a one-quasiparticle configuration
\begin{align}
\ket{\Psi^{N_{odd}}_{i}} = \beta_i^\dagger \ket{\Psi_{0}},
\end{align}
where $\beta_i^\dagger$ is a creation operator of the quasiparticle state $i$. Here the pairing correlation
is taken into account in the framework of the Hartree-Fock-Bogoliubov (HFB) theory, and 
$\ket{\Psi_{0}}$ is the HFB ground state for the neighbor even-$N$ nucleus. 
An initial state of the neutron capture reaction consisting of 
the odd-$N$ nucleus and an impinging neutron is given by
\begin{align}
\ket{\Psi_{i}^{N_{odd}} \otimes \phi_p} = \beta_p^\dagger \beta_i^\dagger\ket{\Psi_{0}}=\ket{\Phi_{i,p}^{N_{even}}}.
\end{align}
Here $\phi_p$ and $\beta_p^\dagger$ represent the scattering neutron with quantum number
$p$ (momentum or partial waves).

Final states of the reaction are low-lying excited states or the ground state of even-$N$ nucleus
with $N_{even}=N_{odd}+1$. We assume that they are the QRPA excited states 
 \begin{align} \label{excited_state_nu}
\ket{\nu L_{\nu}M_{\nu}} =\hat{O}_{\nu L_{\nu}M_{\nu}}^\dagger \ket{\Psi_{0}}
\end{align}
or the ground state
\begin{align}
\ket{0_{\rm gs}^{+}} = \ket{\Psi_{0}},
\end{align}
where $\hat{O}_{\nu L_{\nu}M_{\nu}}^\dagger$ is the QRPA creation operator specified with quantum numbers
$\nu L_{\nu}M_{\nu}$ .

We shall consider an inverse process which is the photoabsorption reaction with initial state $\ket{\nu L_{\nu}M_{\nu}}$ (or the ground state
$\ket{0_{\rm gs}^+}$) of the even-$N$ nucleus,
and final states are excited states decaying via one-neutron emission. The cross section of this  process  
is described in terms of the transition $T$ matrix
\begin{align} \label{T_matrix}
\bra{\Psi_{i,p}(E)} \hat{M}_\lambda \ket{\nu L_{\nu}M_{\nu}},
\end{align}
where $\hat{M}_\lambda$ is an electromagnetic multipole operator and 
$\ket{\Psi_{i,p}(E)}$ is excited states with continuum spectrum, 
located above the neutron separation energy. The continuum excited state
$\ket{\Psi_{i,p}(E)}$ is specified by an asymptotic boundary condition $\ket{\Psi_{i,p}(E)} \sim
\ket{\Phi_{i,p}^{N_{even}}}=\ket{\Psi_{i}^{N_{odd}} \otimes \phi_p}$.
In the present approach, we describe the $T$ matrix, Eq.~(\ref{T_matrix}), by means of the linear response formalism of the cQRPA theory.

\subsection{The photoabsorption cross section and the strength function}
The initial states (the ground state or the low-lying excited states)  of photoabsorption reaction
are QRPA excited states $\ket{\nu L_{\nu} M_{\nu}}$ with excitation energy $E_{\nu}$, or the ground state $\ket{0_{\rm gs}^+}$.
We express excited states populated by the photoabsorption as $\ket{kLM(E)}$.  
Since the excited states are embedded in the continuum, we explicitly write the excitation energy $E$ in addition to 
the angular momentum quantum
numbers $LM$. $k$ is other remaining quantum numbers. We need to specify the continuum excited states $\ket{kLM(E)}$ with the
scattering state $\ket{\Phi_{i,p}^{N_{even}}}=\ket{\Psi_{i}^{N_{odd}} \otimes \phi_p}$ in order to calculate the $(\gamma,n)$ cross section.
The quantum number $k$ is not specified for the moment, and will be defined later (section \ref{Concrete_expressions}).
Normalization is $\langle k'L'M'(E') | kLM(E) \rangle =\delta_{k'L'M', kLM} \delta(E-E')$.

The photoabsorption cross section of the transition from $\ket{\nu L_{\nu}M_{\nu}}$ to states with angular momentum $L$ and
energy $E = E_{\nu} + E_{\gamma}$ is given generally by \cite{Ring1980,Bertulani2004,Thompson2009}
\begin{align}
\sigma^{\lambda}_{\nu L_{\nu} + \gamma \to L}(E_{\gamma}) &= f_\lambda(E_{\gamma}) \sum_{k} B(M_{\lambda}, \nu L_{\nu} \to kL(E))
=\frac{f_\lambda(E_{\gamma}) }{2L_{\nu} + 1} S(M_{\lambda}; \nu L_{\nu}, L; E) 
\end{align}
for electromagnetic multipole $\hat{M}_{\lambda \mu}$ transition with photon energy $E_{\gamma}$ in terms of the reduced matrix element 
\begin{align}
B(M_{\lambda}, \nu L_{\nu} \to kL(E)) &= \frac{1}{2L_{\nu}+1} |\braket{kL(E) ||\hat{M}_{\lambda}|| \nu L_{\nu}}|^{2},
\end{align}
or the strength function
\begin{align} \label{strength_function_general}
S(M_{\lambda}; \nu L_{\nu}, L; E) &= \sum_{kM\mu M_{\nu}} |\bra{kLM(E)} \hat{M}_{\lambda \mu} \ket{\nu L_{\nu}M_{\nu}}|^{2}  \notag \\
&= \sum_{k} |\braket{kL(E) ||\hat{M}_{\lambda}|| \nu L_{\nu}}|^{2} ,
\end{align}
and the kinematical factor
\begin{align}
f_{\lambda}(E_{\gamma}) &= \frac{(2 \pi)^{3} (\lambda + 1)}{\lambda [(2 \lambda + 1) !!]^{2}} \left( \frac{E_{\gamma}}{\hbar c} \right)^{2 \lambda - 1}.
\end{align}

Note that we describe the continuum excited states $\ket{kLM(E)}$ by means of the QRPA, in particular the cQRPA. Namely
we assume that the Hilbert space beyond the QRPA, i.e., that spanned by four- and higher-multiple-quasiparticle configurations
do not play direct roles to the process under consideration. This may be justified for the neutron-rich nuclei close to the drip line, where
the neutron separation energy is low, hence the excitation energy relevant to the low-energy neutron capture reaction is also low.

Note that the low-lying excited state $\ket{\nu L_{\nu}M_{\nu}}$ is given by Eq.~(\ref{excited_state_nu}) using
the QRPA creation operator $\hat{O}^{\dag}_{\nu L_{\nu}M_{\nu}}$, and the continuum excited states
$\ket{kLM(E)}$ are also the QRPA states.  In this case, 
the strength function, Eq.~(\ref{strength_function_general}), 
 is rewritten as \cite{Saito2021}
\begin{align}
\label{strength_function_for_F}
S(M_{\lambda}; \nu L_{\nu}, L; E)   =  \sum_{k} |\braket{kL(E) ||\hat{F}_{L}|| 0^{+}_{\rm gs}}|^{2}  \equiv S(F_{L}; E),
\end{align}
using a newly introduced operator 
\begin{align}
\hat{F}_{L M} \equiv \sum_{\mu M_{\nu}} \langle \lambda \mu L_{\nu} M_{\nu} | L M \rangle [\hat{M}_{\lambda \mu}, \hat{O}^{\dag}_{\nu L_{\nu}M_{\nu}}],
\end{align}
defined as a commutator of the electromagnetic operator $\hat{M}_{\lambda\mu}$ and the
QRPA creation operator $\hat{O}^{\dag}_{\nu L_{\nu}M_{\nu}}$.
Note that the operator $\hat{F}_{LM}$ is a one-body but non-local operator \cite{Saito2021,Saito2023}.

\subsection{Generalized linear response formalism and the $T$ matrix of $(\gamma,n)$ process}

We now explain the linear response formalism to evaluate the strength function $S(F_{L};  E)$ 
and a method to evaluate the $T$ matrix of $(\gamma,n)$ process. It is similar to that described in 
Ref.~\cite{Matsuo2001,Matsuo2015}, but is generalized to treat the non-local operator $\hat{F}_{LM}$.
The formalism is also regarded as an extension of that of Ref.~\cite{Saito2023}.
In this subsection, we omit angular momentum quantum numbers for simplicity.

\subsubsection{Definitions and notations}

The quasiparticle states are single-particle excitations in pair correlated system, and it is 
defined by the HFB equation (the Bogoliubov-de Genne equation),
\begin{align} \label{HFB_equation}
\begin{pmatrix}
\hat{t} + \Gamma_{HF} -\lambda  & \Delta \\ 
\Delta & - \hat{t} - \Gamma_{HF} +\lambda
\end{pmatrix}
\phi_i(\vecrs) = E_i \phi_i(\vecrs),
\end{align}
where $\Gamma_{HF}$ and $\Delta$ are the Hartree-Fock and the pair potentials, respectively, and
\begin{align}
\phi_i(\vecrs) = 
\begin{pmatrix}
\varphi_{i,1}(\vecrs) \\
\varphi_{i,2}(\vecrs) 
\end{pmatrix}
\end{align}
is a two-component wave function of the quasiparticle state $i$ with the excitation energy $E_i$. Here
$\vecrs$ is the coordinate and spin variables $\sigma=\pm \frac{1}{2}$. In the following we use a shorthand
notation $x=\vecrs$. The
creation and annihilation operators $\beta^{\dag}_{i}, \beta_{i}$ of the quasiparticle $i$ are related to the
field operators of creation and annihilation of the nucleon $\psi^{\dag}(x),\psi(x)$ as
\begin{align}
\Psi(x)=
\begin{pmatrix}
\psi(x) \\
\psi^\dag(\tilde{x})
\end{pmatrix}
=\sum_i \left\{\phi_i(x) \beta_i  +   \bar{\phi}_{\tilde{i}}(x)\beta_i^\dag \right\}.
\end{align}
Note the $\Psi(x)$ is a Nambu representation of the field operators, and 
\begin{align}
\bar{\phi}_{\tilde{i}}(x)= 
\begin{pmatrix}
-\varphi_{i,2}^*(\tilde{x}) \\
\varphi_{i,1}^*(\tilde{x}) 
\end{pmatrix}
\end{align}
is a conjugate of $\phi_i(x)$ having the negative energy $-E_i$, where
$\varphi(\tilde{x})=(-2\sigma)\varphi(\vecr -\sigma)$.


A one-body operator $\hat{V}$ which includes pair-addition and pair-removal fields is generally written as
 \begin{align} \label{general_def_V}
\hat{V} &= \iint dxdy \left\{ V_{0}(x,y) \psi^{\dag}(x) \psi(y) + \frac{1}{2} V_{a}(x,y) \psi^{\dag}(x) \psi^{\dag}(\tilde{y}) + \frac{1}{2} V_{r}(x,y) \psi(\tilde{x}) \psi(y) \right\},
 \end{align}
where $\int dx = \sum_{\sigma} \int d \vecr$.

Using the Nambu represetation, the one-body operator $\hat{V}$ is also written as
\begin{align} \label{matrix_form_V}
\hat{V} 
&= \frac{1}{2} \iint dx dy
\Psi^{\dag}(x)
\begin{pmatrix}
V_{0}(x,y) & V_{a}(x,y) \\
V_{r}(x,y) & -V_{0}(\tilde{y},\tilde{x}) \\
\end{pmatrix}
\Psi(y) + const. \notag \\
&= \frac{1}{2} \iint dx dy
\Psi^{\dag}(x) {\cal V}(x,y)
\Psi(y) +const.,
\end{align}
where $ {\cal V}(x,y)$ is a $2 \times 2$ matrix given in the first line. The one-body operator
is expressed also as 
\begin{align} \label{four_term_def_V}
\hat{V}= \frac{1}{2}\iint dx dy \sum_\alpha V_\alpha(x,y) \hat{\rho}_\alpha(y,x)
\end{align}
in terms of the density matrix operators $\hat{\rho}_\alpha(x,y)$ and associated matrix elements
$V_\alpha(x,y)$ and $\hat{\rho}_\alpha(x,y)$, which are  defined by
\begin{align} \label{four_term}
V_{\alpha}(x,y)& =V_0(x,y), V_a(x,y), V_r(x,y), -V_0(\tilde{y},\tilde{x}), \\
\hat{\rho}_{\alpha}(x,y) &= \psi^{\dag}(y) \psi(x), \psi^{\dag}(y) \psi^{\dag}(\tilde{x}),
 \psi(\tilde{y}) \psi(x), \psi(\tilde{y})\psi^\dagger(\tilde{x}),
\end{align}
in the corresponding order.
The index $\alpha$ refers to 11, 12, 21 and 22 components of the $2 \times 2$ matrix ${\cal V}(x,y)$
in Eq.~(\ref{matrix_form_V}). 
Correspondingly ${\cal V}(x,y)$ is  expressed also as
\begin{align}
{\cal V}(x,y)=\sum_{\alpha} {\cal A}_{\alpha} V_{\alpha}(x,y),
\end{align}
with
\begin{align}
{\cal A}_{\alpha} = 
\begin{pmatrix}
1 & 0 \\
0 & 0 \\
\end{pmatrix},
\begin{pmatrix}
0 & 1 \\
0 & 0 \\
\end{pmatrix},
\begin{pmatrix}
0 & 0 \\
1 & 0 \\
\end{pmatrix},
\begin{pmatrix}
0 & 0 \\
0 & 1 \\
\end{pmatrix}.
\end{align}

Now the commutator operator $\hat{F}=[ \hat{M}, \hat{O}^{\dag}_{\nu}]$ can be evaluated as follows.
We first note that the electromagnetic operator $\hat{M}$, assuming for simplicity as a local one-body
operator,  is written as
\begin{align}
\hat{M} &= \int dx f(x) \psi^{\dag}(x) \psi(x)  = \frac{1}{2} \int dx \Psi^{\dag}(x) {\cal M}(x) \Psi(x) + const.,
\end{align}
where 
\begin{align}
{\cal M}(x) =
\begin{pmatrix}
f(x) & 0 \\
0 & -f(\tilde{x}) \\
\end{pmatrix}.
\end{align}
The QRPA creation operator is given in terms of the
quasiparticle operators $\beta^{\dag}_i$ and $\beta_i$ as
\begin{align}\label{XY}
\hat{O}^{\dag}_{\nu}=\sum_{i<j}X_{ij}^\nu \beta_i^\dagger\beta_j^\dagger - Y_{ij}^\nu \beta_j\beta_i . 
\end{align}
The commutator operator $\hat{F}=[ \hat{M}, \hat{O}^{\dag}_{\nu}]$ is then calculated as
\begin{align}
\hat{F} 
&= \frac{1}{2} \iint dx dy
\Psi^{\dag}(x) {\cal F}(x,y)
\Psi(y),
\end{align}
with the matrix elements ${ \cal F}(x,y)$  given by
\begin{align}
{\cal F}(x,y)= 
{\cal M}(x)
\bar{{\cal R}}^{\rm (tr)}_{\nu}(x,y)
-
\bar{{\cal R}}^{\rm (tr)}_{\nu}(x,y)
{\cal M}(y).
\end{align}
Here 
\begin{align}
\label{pseudo_transition_density}
\bar{{\cal R}}^{\rm (tr)}_{\nu}(x,y) 
&= \sum_{ij} X^{\nu}_{ij} \phi_{i}(x) \bar{\phi}^{\dag}_{\tilde{j}}(y) + Y^{\nu}_{ij} \bar{\phi}_{\tilde{i}}(x) \phi^{\dag}_{j}(y)
\end{align}
is expressed in terms of the QRPA amplitudes $X_{ij}^\nu$ and $Y_{ij}^\nu$ of the  low-lying excited state $\ket{\nu}$,
which can be evaluated also in the linear response formalism of the QRPA \cite{Shimoyama2013}.
We call 
$\bar{{\cal R}}^{\rm (tr)}_{\nu}(x,y)$  {\it pseudo transition density matrix} for the low-lying excited state
$\ket{\nu}$ \cite{Saito2021,Saito2023}
(see also Appendix B).

When we consider the strength function for transitions from the ground state, we simply replace $\hat{F}$ by $\hat{M}$.

\subsubsection{Generalized linear response equations}

In the framework of TDHFB or the 
time-dependent density-functional theory (TDDFT), the system is driven by 
the self-consistent nucleon mean field $\hat{U}(t)$, which is a functional of
time-dependent one-body densities.
Relevant densities are
four kinds of density matrices $\rho_{\alpha}(x,y)=\langle \psi^{\dag}(y) \psi(x) \rangle, \langle \psi(\tilde{y}) \psi(x) \rangle,  \langle \psi^{\dag}(y) \psi^{\dag}(\tilde{x}) \rangle, \langle \psi(\tilde{y}) \psi^{\dag}(\tilde{x}) \rangle$.

To describe the strength functions, we consider the linear response of the system against the external perturbing field, which in the present case
is the commutator operator $\hat{F}=[\hat{M}, \hat{O}_\nu^\dagger]$ or the electromagnetic operator $\hat{M}$.
The perturbation causes fluctuations $\delta\rho_{\alpha}(x,y,\omega)$ in one-body density matrices.
Here we express in the frequency domain.
Since the self-consistent nucleon mean field $\hat{U}[\rho_\alpha]$ is a functional of the densities,
the density fluctuations $\delta\rho_\alpha$ induce
fluctutation in the mean-field potential, often called induced potential 
$\delta\hat{U}_{\rm ind}=\sum_\alpha \frac{\del \hat{U}}{\del \rho_\alpha}\delta\rho_\alpha$.

Thus a 
net perturbing field is $\hat{F} + \delta \hat{U}_{\rm ind}(\omega)\equiv \hat{V}_{\rm scf}(\omega)$, consisting of 
the external perturbing field $\hat{F}$ and
the induced field $\delta \hat{U}_{\rm ind}(\omega)$. As we describe in Appendix A,
the density fluctuations are governed by the linear response equation
\begin{align} \label{density_matrix_response}
\delta \rho_{\alpha}(x,y,\omega) &= \iint dx^{'} dy^{'} \sum_{\beta} R^{\alpha \beta}_{0}(x, y; y^{'}, x^{'}; \omega) 
\left[ F_{\beta}(x^{'},y^{'}) + \delta U^{\rm ind}_{\beta}(x^{'},y^{'},\omega) \right].
\end{align}
Here $F_{\beta}(x^{'},y^{'})+\delta {U}_\beta^{\rm ind}(x^{'},y^{'},\omega)$ are matrix elements of
$\hat{V}_{\rm scf}(\omega) =\hat{F}+\delta \hat{U}_{\rm ind}(\omega)$ in the notation defined by Eqs.~(\ref{four_term_def_V}) and (\ref{four_term}).

The function $R^{\alpha \beta}_{0}(x,y;y^{'},x^{'};\omega)$ is  an unperturbed response function for the density matrix
defined by Eqs.~(\ref{R_0_Green}) and (\ref{R_0_continuum}) in Appendix A.
It is expressed also in the spectral representation as
\begin{align} \label{R_0_spectrum}
R^{\alpha \beta}_{0}(x,y;y^{'},x^{'};\omega)&= \frac{1}{4                                                                                                                                                                                                                                                                                                                                                                                                                                              } \sum_{ij} \Big\{ \bra{0} \hat{\rho}_{\alpha}(x,y) \ket{ij} \bra{ij} \hat{\rho}_{\beta}(y^{'},x^{'}) \ket{0} \frac{1}{\hbar \omega + i \epsilon - E_{i} - E_{j}} \notag \\
&- \bra{0} \hat{\rho}_{\beta}(y^{'},x^{'}) \ket{ij} \bra{ij} \hat{\rho}_{\alpha}(x,y) \ket{0} \frac{1}{\hbar \omega + i \epsilon + E_{i} + E_{j}} \Big\} ,
\end{align}
where $\ket{0}$ denotes the HFB ground state $\ket{\Psi_{0}}$ and $\ket{ij}=\beta^{\dag}_{i} \beta^{\dag}_{j} \ket{0}$ is a two-quasiparticle configuration state.
$\epsilon$ is a positive infinitesimal constant.

Solving the linear response equation, we obtain the density matrix responses $\delta \rho_{\alpha}(x,y,\omega)$ 
as well as the strength function
\begin{align} \label{strength_function_for_F_2}
S(F; \hbar\omega)&=\sum_k |\bra{k} \hat{F} \ket{0}|^2 \delta(\hbar\omega-E_{k}) \notag \\
&=-\frac{1}{2\pi}{\rm Im} \iint dx dy \sum_{\alpha} F_\alpha^\dagger(y,x)\delta\rho_\alpha(x,y,\omega),
\end{align}
where $F_\alpha^\dagger(y,x) =F_0^*(x,y), F_r^*(x,y), F_a^*(x,y), -F_0^*(\tilde{y},\tilde{x})$.

\subsubsection{Zangwill-Soven decomposition and $T$ matrix of $(\gamma, n)$ process} 

We shall evaluate partial $(\gamma,n)$ cross sections for individual decay channels. Following  Zangwill and Soven \cite{Zangwill1980},
the strength function Eq.~(\ref{strength_function_for_F_2}) is rewritten as 
 \begin{align}
\label{strength_function_vscf}
S(F;  \hbar \omega) &= -\frac{1}{2\pi} {\rm Im} \iiiint dx dy dx^{'} dy^{'} \sum_{\alpha \beta} {V}^{\rm scf \dag}_{\alpha}(y,x,\omega) R^{\alpha \beta}_{0}(x,y;y^{'},x^{'};\omega) V^{\rm scf}_{\beta}(x^{'},y^{'},\omega) \\
\label{strength_function_vscf_braket}
&= \sum_{i>j} |\bra{ij} \hat{V}_{\rm scf}(\omega) \ket{0}|^{2} \delta(\hbar \omega - (E_{i} + E_{j})).
\end{align}
Here the matrix element $|\bra{ij} \hat{V}_{\rm scf}(\omega) \ket{0}|^{2}$ is interpreted as a transition probability
to the two-quasiparticle state $\ket{ij}$ (apart from a factor) at excitation energy $\hbar\omega$.

It is noted here that the model space of the cQRPA consists of two-quasiparticle configurations.
Here a quasiparticle is either bound if its wave function is localized around the nucleus, 
or unbound if it becomes scattering wave extending far outside the nucleus. It is distinguished in terms of
 relation between the excitation energy $E_i$ of the quasiparticle state and the Fermi energy $\lambda$: bound for $E_i < |\lambda|$, and unbound for
  $E_i > |\lambda|$ \cite{Dobaczewski1984}. Let us denote
  $n,m$ for bound states, $p(E_{p}), q(E_{q})$ for unbound states with energy $E_{p(q)}$ while $i,j$ cover both. 
  Similarly  the model space of the QRPA, the two-quasiparticle configurations $\ket{ij}=\beta_i^\dagger \beta_j^\dagger\ket{0}$,
  are classified into three categories. 
  i) Both are bound $\ket{mn}$. ii) One is bound while the other is unbound 
  $\ket{np(E_{p})}$. In this case one nucleon is in a scattering state, represented by $p(E_{p})$.
   iii) Both are unbound  $\ket{p(E_{p})q(E_{q})}$, which represents a configuration where 
   two nucleons are in scattering states $p(E_{p})$ and $q(E_{q})$. 
   
\begin{figure}[H]
\centering
\begin{minipage}{0.30\columnwidth}
\includegraphics[width=\columnwidth]
{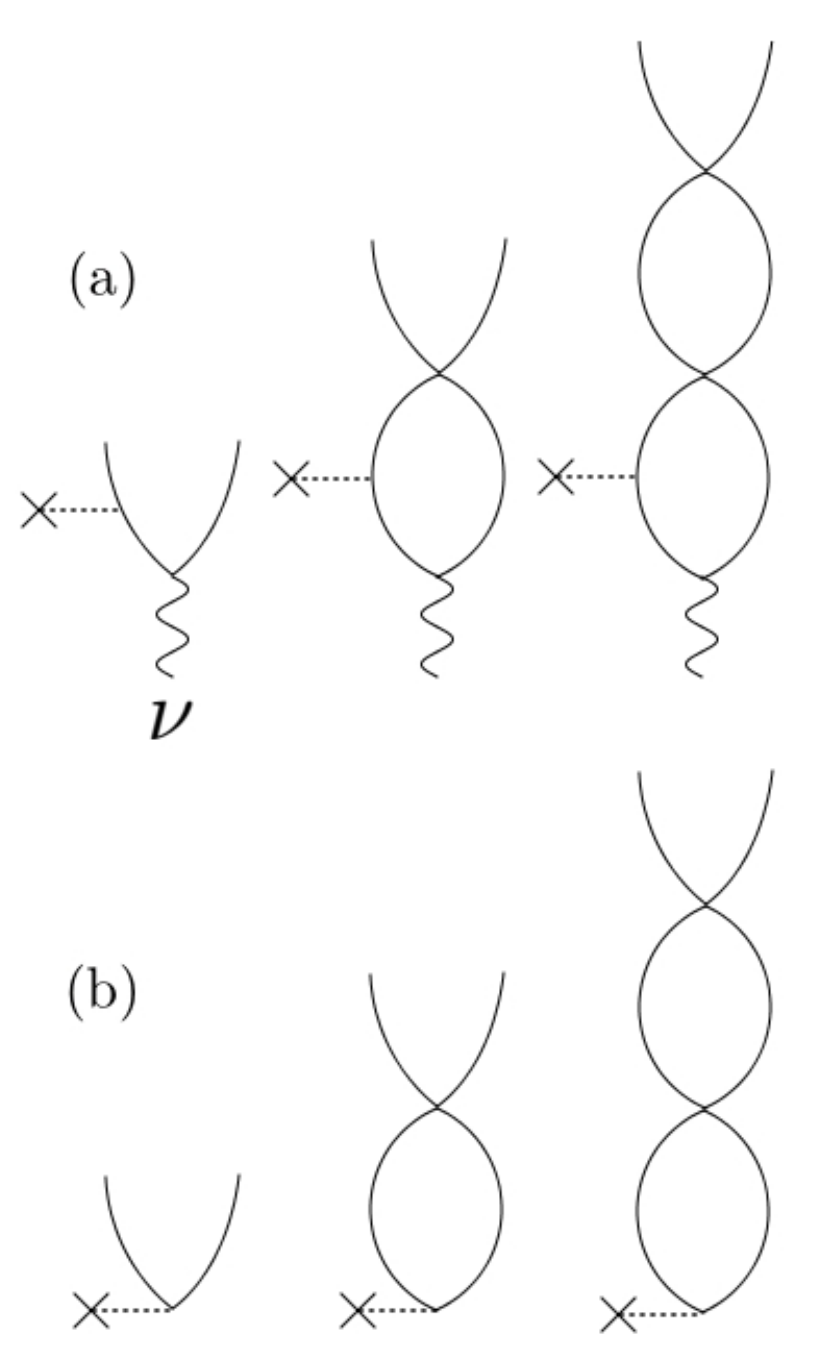}
\end{minipage} 
\caption{(a) The diagrams corresponding to $T$ matrix, Eq.~(\ref{T-matrix}), of the $(\gamma,n)$ reaction
for the initial state $\ket{\nu}$ and the final state consisting one-quasiparticle configuration $n$ and
the scattering neutron $p(E_{p})$.
(b) Same as (a) but the case where the initial state is the ground state $\ket{0^{+}_{\rm gs}}$.}
\label{diagram_theory}
\end{figure}

Consequently the strength function describing the $(\gamma,n)$ reaction 
 corresponds to the category ii): 
\begin{align} \label{strength_function_1c}
S_{1c}(\hbar\omega)&= \sum_n \SumInt_p  |\bra{np(E_p)} \hat{V}_{{\rm scf}}(\omega) \ket{0}|^{2} \delta(\hbar \omega - (E_{n} + E_{p})) \\
&= -\sum_n \frac{1}{\pi}{\rm Im}\iiiint dxdydx'dy' \bar{\phi}^{\dag}_{\tilde{n}}(y)  {\cal V}^{\dag}_{\rm scf}(y,x,\omega) {\cal G}_{0c}(x, x^{'}, - E_{n} + \hbar \omega + i \epsilon)
 {\cal V}_{{\rm scf}}(x^{'},y^{'},\omega) \bar{\phi}_{\tilde{n}}(y^{'}),
\end{align}
with ${\cal V}^{\dag}_{\rm scf}(y,x,\omega)=(({\cal V}_{\rm scf}(x,y,\omega))^*)^T$. 
Here, $\SumInt_p$ denotes a summation over continuum quasiparticle states.
${\cal G}_{0c}$ is the continuum part of the quasiparticle Green's function ${\cal G}_{0}$ defined in Appendix A, subtracting contributions of bound quasiparticle states.
Note that each term of Eq.~(\ref{strength_function_1c}), 
\begin{align}\label{T-matrix}
\bra{np(E_p)} \hat{V}_{\rm scf}(\omega) \ket{0} = \bra{np(E_p)} \hat{F} + \delta\hat{U}_{\rm ind}(\omega) \ket{0},
\end{align}
is the $T$ matrix of the $(\gamma,n)$ reaction with specified configuration of the final scattering state, i.e., one-quasiparticle  state 
$n$ and scattering neutron with quantum number $p(E_{p})$ \cite{Saito2023}. 
It can be represented  diagrammatically as Fig.~\ref{diagram_theory}.
The partial photoabsorption cross section is then obtained by multiplying the kinematical factor
 $f_\lambda(E_\gamma)$. 
 
\subsection{Concrete expressions for $(\gamma ,n)$ and $(n, \gamma)$ cross sections} \label{Concrete_expressions}
We perform numerical calculations under the assumption of spherical symmetry of the system, and describe
 the response with partial waves of the single-particle (quasiparticle) states and the multipolarity $LM$ of the 
 total system.  
 We specify the quasiparticle configuration of the odd-$N$ nucleus with $nlj$, and the escaping neutron with kinetic energy $\epsilon_{\rm kin}$ (or 
 quasiparticle energy $E_{p}=|\lambda| + \epsilon_{\rm kin}$) and the partial wave quantum number $l^{'}j^{'}$. The final state $\ket{np(E_{p})}$ is $\ket{[nlj,\epsilon_{\rm kin} l^{'}j^{'}]_{LM}} = \sum_{mm^{'}} \braket{jmj^{'}m^{'} | LM} \beta^{\dag}_{nljm} \beta^{\dag}_{E_{p}l^{'}j^{'}m^{'}} \ket{0}$ with $L$ being the total angular momentum. 
 Consequently we obtain the expression for the partial photoabsorption cross section for the specific channel of neutron decay as
\begin{align}
\label{gamma_n_cross_section_radial}
\sigma^{\lambda}_{\nu L_{\nu} + \gamma \to [nlj, \epsilon_{\rm kin} l^{'}j^{'}]_{L}}(E_{\gamma}) 
&= - \frac{f_{\lambda}(E_{\gamma})}{\pi(2L_{\nu} + 1)} {\rm Im} \iiiint dr_{x} dr_{y} dr_{x^{'}} dr_{y^{'}} \notag \\
&\times \bar{\phi}^{T}_{nlj}(r_{y}) {\cal V}^{\rm scf \dag}_{L, l^{'}j^{'}, lj}(r_{y}, r_{x}, \omega) {\cal G}_{0c, l^{'}j^{'}}(r_{x}, r_{x^{'}}, -E_{nlj} + \hbar \omega + i \epsilon) {\cal V}^{\rm scf}_{L, l^{'}j^{'}, lj}(r_{x^{'}}, r_{y^{'}}, \omega) \bar{\phi}_{nlj}(r_{y^{'}}),
\end{align}
with
\begin{align}
 {\cal V}^{\rm scf}_{L, l^{'}j^{'}, lj}(r_{x}, r_{y}, \omega) = {\cal F}_{L, l^{'}j^{'}, lj}(r_{x}, r_{y}) 
 + \langle l^{'}j^{'} || Y_{L}|| lj \rangle \sum_\beta \frac{\delta {\cal U}}{\delta \rho_\beta}(r_{x}) \frac{1}{r^{2}_{x}} \delta \rho_{\beta L}(r_{x}, \omega) \delta(r_{x} - r_{y}).
\end{align}
Here $ {\cal V}^{\rm scf}_{L, l^{'}j^{'}, lj}(r_{x}, r_{y}, \omega)$ and $ {\cal F}_{L, l^{'}j^{'}, lj}(r_{x}, r_{y}) $ are matrix elements
of $\hat{V}_{\rm scf}$ and $\hat{F}_L$ respectively (cf.~Appendix B).
$\hbar \omega = E_{\gamma} + E_{\nu}$ is the excitation energy of the states populated by the photoabsorption. 
The energy of emitted neutron is $\epsilon_{\rm kin}=-E_{nlj} + \hbar \omega - |\lambda|$.

Finally, using detailed balance we obtain the radiative neutron capture cross section for the reaction $\ket{ [ \Psi^{N_{odd}}_{nlj} \otimes \phi_{\epsilon_{\rm kin} l^{'}j^{'}} ]_{L}} \to \ket{\nu L_{\nu}} + \gamma$:
\begin{align}
\sigma^{\lambda}_{[nlj,\epsilon_{\rm kin} l^{'}j^{'}]_{L} \to \nu L_{\nu} + \gamma}(\epsilon_{\rm kin}) = \frac{2L_{\nu} + 1}{2j + 1} \frac{E_{\gamma}}{2mc^{2} \epsilon_{\rm kin}} \sigma^{\lambda}_{\nu L_{\nu} + \gamma \to [nlj,\epsilon_{\rm kin} l^{'}j^{'}]_{L}}(E_{\gamma})
\end{align}
for an incident neutron with partial wave $l^{'}j^{'}$ and energy $\epsilon_{\rm kin}$ colliding on odd-$N$ nucleus with one-quasiparticle configuration $\beta^{\dag}_{nlj}\ket{0}$, decaying to the excited state $\ket{\nu L_{\nu} M_{\nu}}$.
The total cross section is obtained by summing all contributions of the partial waves $l^{'}j^{'}$ and the total angular momentum $L$ of the system.

\section{numerical examples} \label{numerical_examples}
We shall demonstrate new features of 
the present theory with numerical calculations with a focus on roles of low-lying quadrupole and octupole collectivities
as well as the pairing correlation. 
 We shall discuss two examples,
$^{91}{\rm Zn} (n,\gamma)^{92}{\rm Zn} $ and $^{89}{\rm Ge} (n,\gamma){}^{90}{\rm Ge} $. We have chosen these
examples for the following reasons. First, these nuclei have
small neutron separation energy of the order of $S_{1n} \sim 2-3$ MeV, and 
are near the $N=50$ waiting point on the $r$-process path and this area is of interest in terms of the weak $r$ process \cite{Surman2014}.
Second, the target odd-$N$ nuclei $^{91}{\rm Zn}$ and $^{89}{\rm Ge}$ are expected to have
the valence neutron in the $2d_{3/2}$ orbit, and this situation 
 is favorable for our discussion. As we show below, the QRPA calculation shows presence of
many resonances with $L^\pi=2^+$ around the threshold energy, and
a captured neutron in the $s$-wave combined with the valence neutron in the $d$-orbit will produce
these resonances. Third, the QRPA calculation predicts low-lying collective $3_1^{-}$ state  in
$^{92}{\rm Zn}$ and $^{90}{\rm Ge}$. One may expect impact of the collectivity
for the $E1$ transitions from the $2^+$ resonances to the  $3^{-}_{1}$ state. 
 Note also that these nuclei are expected to be
spherical or only weakly deformed ($\beta_2 <0.2$) in the ground state as many mean-field calculations predict \cite{Massexpl}.
 We have performed calculations also for other isotopes in this mass region,
and we focus below the cases with clear characteristic features.

\subsection{setting}
We use the Skyrme energy density functional model and the effective pairing interaction of the contact type to construct the self-consistent mean field, the HFB ground state, the residual interaction, and the excited states. 
The adopted Skyrme parameter sets are SkM$^{*}$ \cite{Bartel1982} for $^{92}{\rm Zn}$ and SLy4 \cite{Shabanat1998} for $^{90}{\rm Ge}$, and the density-dependent delta interaction (DDDI),
\begin{align}
v_{\rm pair,\tau}(\mathbf{r},\mathbf{r}^{'})=v_{0} \frac{1-P_{\sigma}}{2} \left[ 1 - \eta \left( \frac{\rho_{\tau}(\mathbf{r})}{\rho_{c}} \right)^{\alpha} \right] \delta(\mathbf{r} - \mathbf{r}^{'}),
\end{align}
with $v_{0}=-458.4$ MeV, $\rho_{c}=0.08$ fm$^{-3}$ are used \cite{Matsuo2010}.
The subscript $\tau$ indicates neutron or proton component.
The cutoff energy of quasiparticle is $E_{\rm cut}=60$ MeV and the cutoff orbital angular momentum is $l_{\rm cut}=8$. 
The radial HFB equation is solved in a spherical box $r < R_{\rm max}$ with mesh $\Delta r=0.2$ fm in the box boundary condition $\phi(r=R_{\rm max})=0$ with $R_{\rm max}=20$ fm.
The DDDI dimensionless parameter sets are fitted to the experimental pairing gaps of stable Zn and Ge isotopes;  
 $\alpha=0.59,\eta=0.76$ for $^{92}{\rm Zn}$ and $\alpha=0.59,\eta=0.69$ for $^{90}{\rm Ge}$ with $\alpha$ given in Ref.~\cite{Matsuo2010}.

The setting for the QRPA calculation is as follows. We adopt the Landau-Migdal approximation for the residual interaction
in the linear response equation. Namely the residual particle-hole interaction is given by a contact force with
the Landau-Migdal parameters $F_{0}(\mathbf{r}), F_{0}^{'}(\mathbf{r})$, for which we use local density approximation.
 Under this approximation, we solve the linear response equation to obtain the fluctuations in the local density and the local pair density, and the 
corresponding local particle-hole and  local pair fields for the induced field. As is often done in many works adopting the Landau-Migdal
approximation, the residual interaction is multiplied by a renormalization factor $f$ so that the spurious dipole mode
associated with the displacement motion appears at zero excitation energy. There is no experimental information on excited states
of  $^{92}{\rm Zn}$ and $^{90}{\rm Ge}$, which are the final states of the $\gamma$ transition. The lowest-lying $2^+$ state is observed
at excitation energy 0.599 MeV in a neighbor nucleus $^{84}{\rm Zn}$ \cite{Shand2017},
and at 0.527 MeV in $^{86}{\rm Ge}$ \cite{Miernik2013}. In calculating the quadruple mode, we use a renormalization factor 
which reproduces these excitation energies. For other multipoles, we use the same value of $f$ as that of the dipole mode.

In order to obtain the density fluctuations and the induced field
in the continuum spectrum above the neutron separation energy, we use the response function Eq.~(\ref{R_0_continuum}) in the
Green's function representation. Here 
the quasiparticle wave functions constructing Green's function are connected to the Hankel functions at $R_{\rm max}$ 
in order to treat the unbound quasiparticle states with proper boundary conditions. The small imaginary constant
$i\epsilon$ in the response and Green's functions is set as $\epsilon=0.1$ MeV otherwise stated.
The constant $\epsilon$ in the Green's function in  Eq.~(\ref{gamma_n_cross_section_radial}) for the $(\gamma,n)$ cross section 
is taken a very small value $\epsilon=10^{-8}$ MeV in order to calculate the scattering quasiparticle states with precise energy.
The $\gamma$ transition is assumed to be $E1$ type.

\subsection{$^{89}{\rm Ge}(n,\gamma)^{90}{\rm Ge}$}


Quasiparticle states for $^{90}{\rm Ge}$ around the Fermi energy, obtained by
 the HFB calculation using the Skyrme SLy4 functional, are listed in Table \ref{spe_90Ge}. 
The neutron Fermi energy is $\lambda_{n}=-2.14$ MeV, and the bound quasineutron states are those associated with the orbits
$3s_{1/2}$, $2d_{3/2}$, $2d_{5/2}$ and $1g_{7/2}$.  The ground state of the neighbor odd-$N$ nucleus $^{89}{\rm Ge}$ is expected
to have one-quasineutron configuration $3s_{1/2}$ or $2d_{3/2}$, which we denote 
$^{89}{\rm Ge}(1/2^{+})$ or $^{89}{\rm Ge}(3/2^{+})$ in the following. The neutron separation energy of $^{90}{\rm Ge}$ for channels decaying to $^{89}{\rm Ge}(1/2^{+})$ or $^{89}{\rm Ge}(3/2^{+})$
are then $S_{1n}(3s_{1/2})=E_{3s_{1/2}}+|\lambda_{n}|=2.84$ MeV, $S_{1n}(2d_{3/2})=E_{2d_{3/2}}+|\lambda_{n}|=3.31$ MeV, respectively. 
We calculate two cases of the $(n,\gamma)$ cross section, where the target state is $^{89}{\rm Ge}(1/2^{+})$ or $^{89}{\rm Ge}(3/2^{+})$.

\begin{figure}[H]
\centering
\begin{minipage}{0.49\columnwidth}
\includegraphics[width=\columnwidth]
{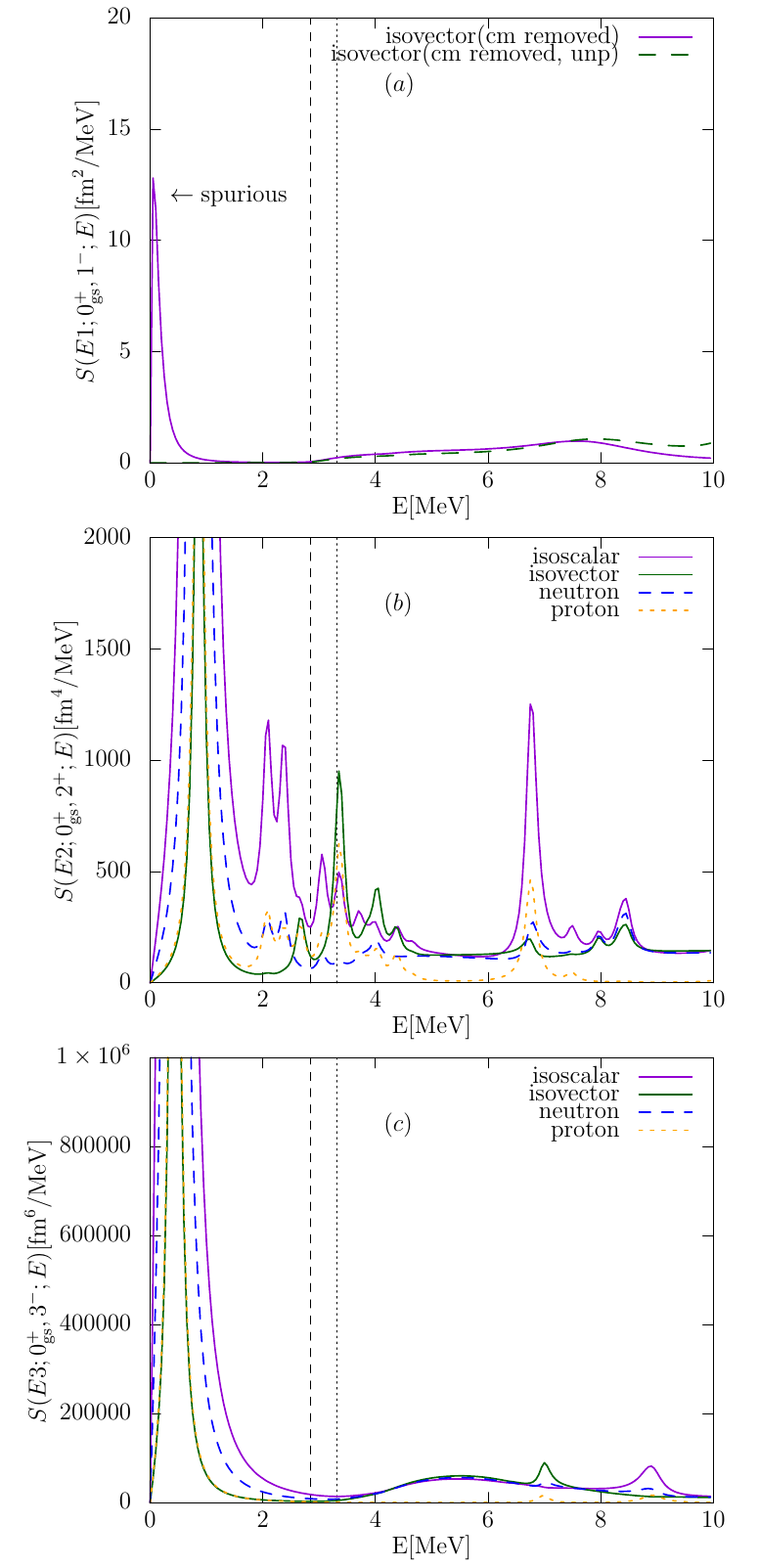}
\end{minipage}  
\caption{(a) The strength function for $L^{\pi}=1^{-}$ excited states in $^{90}{\rm Ge}$, calculated with $\epsilon=0.1$ MeV.
The horizontal axis is the excitation energy of the $1^{-}$ states.
The vertical dashed  and dotted lines indicate one-neutron separation energies $S_{1n}(3s_{1/2})=2.84$ MeV and $S_{1n}(2d_{3/2})=3.31$ MeV,
respectively. (b) and (c) Same as (a) but for $L^{\pi}=2^{+}$ and $3^{-}$ states, respectively.}
\label{strength_function_90Ge}
\end{figure}

Figure \ref{strength_function_90Ge} shows the QRPA strength functions for the dipole, quadrupole and octupole excitations ($L^\pi=1^{-},2^{+},3^{-}$)
of  $^{90}{\rm Ge}$  for isoscalar, isovector, neutron and proton multipole operators. For the dipole ($L^\pi=1^{-}$), the isovector strength  
forms a broad continuous bump above the neutron separation energy, which may be regarded as soft dipole excitation. The peak at zero
energy is the spurious displacement mode.  For the quadrupole excitation ($L^\pi=2^{+}$), there exist many peaks in the energy region shown here.
The lowest peak, corresponding to the $2^{+}_{1}$ state of $^{90}{\rm Ge}$ has excitation energy $E_{2^{+}_{1}}=0.880$ MeV and large
isoscalar quadrupole strength, indicating a clear collective character of surface vibration. The collective nature is seen also in
the forward and backward amplitudes ${X_{ij},Y_{ij}}$, listed in Table \ref{amp_XY_1st2+_90Ge}. We also note that
that there exist peaks  above  the neutron separation energy, indicating presence of resonance states with
$L^\pi=2^{+}$. We shall discuss the role of the resonances in the $(n,\gamma)$ reaction in the following. 
The octupole strength function shows a distinct low-lying peak with peak energy $E=0.881$ MeV. It has dominant isoscalar
character indicating surface octupole vibration. The forward and backward amplitudes ${X_{ij},Y_{ij}}$ of this $3^{-}_{1}$ state
is listed in Table  \ref{amp_XY_1st3-_90Ge}.  In the following we evaluate the cross sections for $^{89}{\rm Ge}(n,\gamma)^{90}{\rm Ge}$,
where the final states of the $\gamma$ transition are the ground state, the $2^{+}_{1}$ state and the $3^{-}_{1}$ state of $^{90}{\rm Ge}$.

\begin{table}[H]
\centering
\caption{Single-quasiparticle energies of the adopted Skyrme energy density functional SLy4 for $^{90}{\rm Ge}$. 
Quasiparticle states corresponding to bound Hartree-Fock orbits around the Fermi energy ($\lambda_{n}= -2.14, \lambda_{p}=-16.96$ MeV) are listed.
Here the quasiparticle state $1g_{7/2}$, for instance, is the one which corresponds to the bound Hartree-Fock orbit $1g_{7/2}$.}
\label{spe_90Ge}
\begin{tabular}{cc p{5mm} cc} \\ \hline \hline
neutrons & $E_i$ (MeV) && protons & $E_i$ (MeV) \\ \hline
$1g_{7/2}$  & 2.02 && $1f_{7/2}$ & 5.40 \\
$2d_{5/2}$ & 1.96 && $2p_{1/2}$ & 2.69 \\
$2d_{3/2}$ & 1.17 && $1f_{5/2}$ & 2.11 \\
$3s_{1/2}$ & 0.70 && $2p_{3/2}$ & 1.80 \\
\hline \hline
\end{tabular}
\end{table}

\begin{table}[H]
\centering
\caption{The QRPA forward and backward amplitudes $X_{ij},Y_{ij}$ of the $2^{+}_{1}$ state of $^{90}{\rm Ge}$. 
Two-quasiparticle configurations with large amplitude $|X_{ij}|>0.1$ are listed.
The QRPA backward and forward amplitudes $X_{ij},Y_{ij}$ are calculated with use of the method of Ref.~\cite{Shimoyama2013}
and discretized quasiparticle states obtained in the box boundary condition.}
\label{amp_XY_1st2+_90Ge}
\begin{tabular}{crr p{5mm} crr} \\ \hline \hline 
neutron config. & \multicolumn{1}{c}{$X_{ij}$} & \multicolumn{1}{c}{$Y_{ij}$} && proton config. & \multicolumn{1}{c}{$X_{ij}$} & \multicolumn{1}{c}{$Y_{ij}$} \\ \hline
$[2d_{3/2}][3s_{1/2}]$ & -0.63 & -0.17 && $[2p_{3/2}][2p_{1/2}]$ & -0.21 & -0.12 \\
$[1g_{7/2}][2d_{3/2}]$ & -0.38 & 0.15 && $[2p_{3/2}]^{2}$ & -0.21 & 0.11 \\
$[2d_{5/2}][3s_{1/2}]$ & -0.31 & 0.099 && $[1f_{5/2}]^{2}$ & -0.20 & 0.11 \\
$[2d_{3/2}]^{2}$ & 0.29 & -0.087 && $[1f_{5/2}][2p_{1/2}]$ & 0.20 & -0.12 \\
$[1g_{7/2}]^{2}$ & 0.22 & -0.091 && $[1f_{5/2}][2p_{3/2}]$ & -0.16 & -0.087 \\
$[2d_{5/2}][2d_{3/2}]$ & -0.21 & -0.10 && $[1f_{7/2}][2p_{3/2}]$ & -0.14 & 0.096 \\
$[1g_{7/2}][2d_{5/2}]$ & -0.13 & -0.077 && & & \\
\hline \hline
\end{tabular}
\end{table}

\begin{table}[H]
\centering
\caption{Same as Table \ref{amp_XY_1st2+_90Ge} but for the $3^{-}_{1}$ state of $^{90}{\rm Ge}$.
Two-quasiparticle configurations with the largest 10 amplitudes are listed. The label $c_1$ in
$c_{1}h_{11/2}$ for example
represents a continuum but discretized unbound quasiparticle state in the partial wave $h_{11/2}$.
The numbering in $c_{i}$ is
given according to the order of the quasiparticle energies for each partial wave.}
\label{amp_XY_1st3-_90Ge}
\begin{tabular}{crr p{5mm} crr} \\ \hline \hline 
neutron config. & \multicolumn{1}{c}{$X_{ij}$} & \multicolumn{1}{c}{$Y_{ij}$} && proton config. & \multicolumn{1}{c}{$X_{ij}$} & \multicolumn{1}{c}{$Y_{ij}$} \\ \hline
$[c_{1}h_{11/2}][2d_{5/2}]$ & 0.88 & -0.66 && $[1g_{9/2}][2p_{3/2}]$ & 0.64 & -0.49 \\
$[1g_{9/2}][c_{1}h_{11/2}]$ & 0.53 & -0.45 && $[1g_{9/2}][1f_{7/2}]$ & 0.42 & -0.35 \\
$[2p_{1/2}][1g_{7/2}]$ & 0.29 & -0.25 && $[1g_{9/2}][1f_{5/2}]$ & -0.19 & -0.15 \\
$[1f_{5/2}][1g_{7/2}]$ & -0.26 & 0.23 && $[c_{2}i_{13/2}][1f_{7/2}]$ & -0.19 & 0.17 \\
$[c_{2}f_{7/2}][2d_{5/2}]$ & -0.18 & 0.15 && $[2d_{3/2}][2p_{3/2}]$ & -0.18 & -0.16 \\
$[c_{2}f_{7/2}][3s_{1/2}]$ & 0.18 & -0.14 && $[1f_{7/2}][2d_{5/2}]$ & -0.18 & 0.16 \\
$[2p_{3/2}][2d_{3/2}]$ & -0.18 & -0.15 && $[2d_{5/2}][2p_{3/2}]$ & -0.17 & 0.15 \\
$[c_{4}j_{15/2}][1g_{9/2}]$ & -0.16 & 0.15 && $[1g_{7/2}][1f_{5/2}]$ & -0.17 & 0.15 \\
$[1f_{7/2}][c_{3}i_{13/2}]$ & -0.15 & 0.14 && $[1f_{5/2}][2s_{1/2}]$ & -0.16 & -0.14 \\ 
$[1f_{5/2}][2d_{3/2}]$ & 0.15 & -0.13 && $[1d_{5/2}][1h_{11/2}]$ & -0.15 & 0.14 \\
\hline \hline
\end{tabular}
\end{table}

\begin{figure}[H]
\centering
\begin{minipage}{0.49\columnwidth}
\includegraphics[width=\columnwidth]
{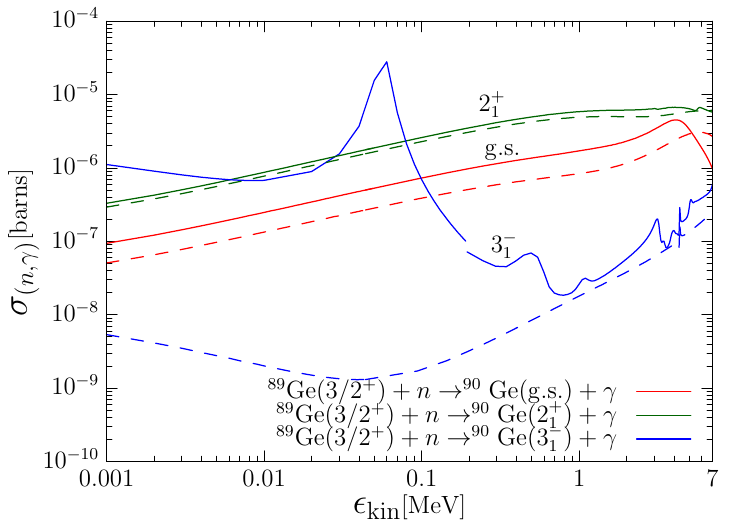}
\end{minipage} 
\caption{The calculated $(n,\gamma)$ cross sections for $^{89}{\rm Ge}(3/2^{+})  +n \to ^{90}{\rm Ge} + \gamma$ with $E1$ transitions populating different low-lying states in $^{90}{\rm Ge}$: the ground state, the quadrupole vibrational state $2^{+}_{1}$, and the octupole vibrational state $3^{-}_{1}$. The solid curves represent the results of the present theory while the dashed curves do the results of the
semi-QRPA approximation, see text. The horizontal axis is kinetic energy $\epsilon_{\rm kin}$ of the incident neutron.
The smoothing parameter $\epsilon$ is $0.1$ MeV, but  smaller smoothing parameter $\epsilon=0.005$ MeV  is used in low-energy region $\epsilon_{\rm kin} < 0.2$ MeV for the transition to the $3^{-}_{1}$ state.}
\label{n_cap_total_cross_section_90Ge_3/2+}
\end{figure}

First we discuss the calculated $(n,\gamma)$ cross section for the $J^\pi=3/2^{+}$ configuration of $^{89}{\rm Ge}$ in the initial channel.
Solid curves in Fig.~\ref{n_cap_total_cross_section_90Ge_3/2+} represent the cross section for the $E1$ transition to
the ground state, the $2^{+}_{1}$ and the $3^{-}_{1}$ states of $^{90}{\rm Ge}$, i.e., for the reaction
$^{89}{\rm Ge}(3/2^{+}) + n \to ^{90}{\rm Ge}({\rm g.s.}, 2^{+}_{1}, 3^{-}_{1}) + \gamma$.

The dotted curves show another calculation in which the QRPA correlation
is neglected in the excited continuum states (scattering states) while the final states of the $\gamma$
transition, i.e., the low-lying $2^{+}_{1}$ and $3^{-}_{1}$ in the present case, are constructed with
the QRPA. Practically we drop off  the induced field ${\cal V}_{\rm ind}$ in Eq.~(\ref{gamma_n_cross_section_radial})  in the calculation. In this case, the initial state with scattering neutron is approximated as
an independent unbound quasineutron state, from which the
$\gamma$ transition occurs directly. Hereafter we call this approximated calculation 
a semi-QRPA calculation since
the QRPA correlation is taken into account only in the low-lying final states.

It is seen that the cross section for the transition to the $2^{+}_{1}$ state is larger than that to the ground state. 
The cross section to the $3^{-}_{1}$ state is the largest at low energy $\epsilon_{\rm kin} \lesim 0.1$ MeV. 
It is noted that there would be no $3^{-}$ states (no negative parity states) below the threshold energy if there is no QRPA correlation.

\begin{figure}[H]
\centering
\begin{minipage}{0.40\columnwidth}
\includegraphics[width=\columnwidth]
{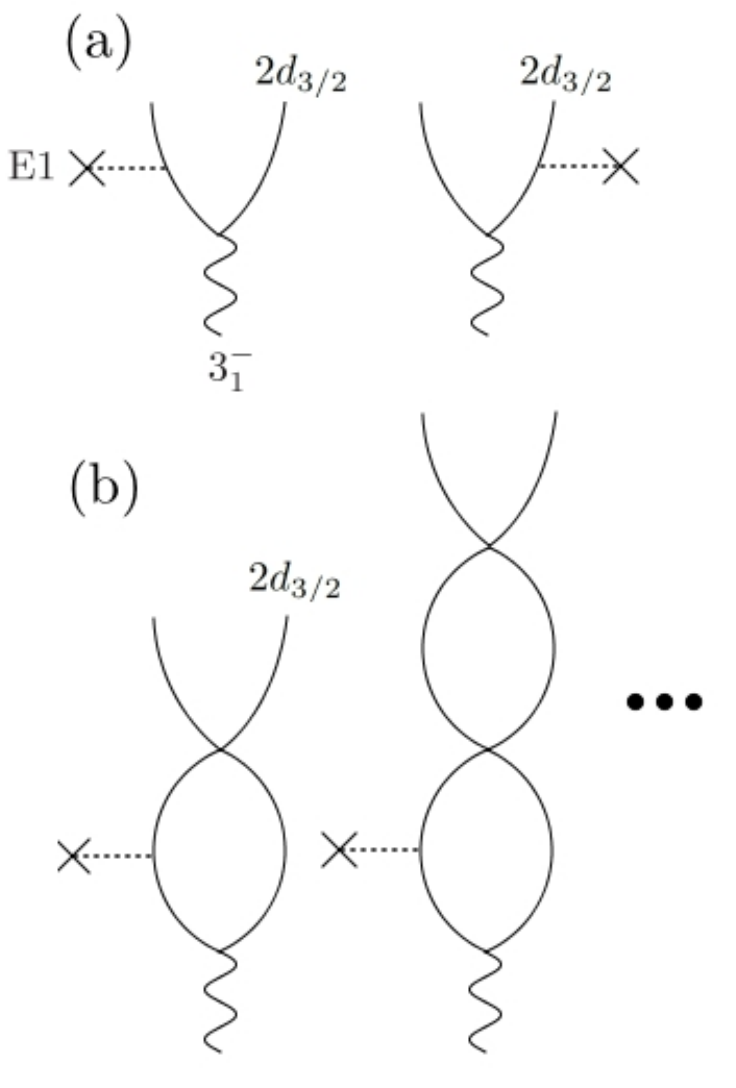}
\end{minipage} 
\caption{(a) Diagrams representing the $T$ matrix in the semi-QRPA approximation, where 
$^{89}{\rm Ge}$ has a one-quasineutron $2d_{3/2}$ configuration and the final
state of the $\gamma$ transition is the $3^{-}_{1}$ state of $^{90}{\rm Ge}$
(b) Diagrams representing the QRPA correlation in the  $T$ matrix. }
\label{diagram_90Ge}
\end{figure}

Distinctive features are seen in the cross section for the transition to the $3^{-}_{1}$ state.
 It shows a very clear resonant behavior around $\epsilon_{\rm kin} \sim 0.03-0.07$ MeV in addition to many other (less significant) resonant peaks at higher energy $\epsilon_{\rm kin} \gesim 0.3$ MeV.
The cross section at this resonance $\epsilon_{\rm kin} \sim 0.03-0.07$ MeV is much larger than the
cross sections for the ground state and for the $2^{+}_{1}$
final states by more than factor of $10^2$ and $10^1$. As seen in the comparison with the semi-QRPA result (dashed line),
the resonance structure originates from the QRPA correlation (Fig.~\ref{diagram_90Ge}(b)) in the continuum excited states 
in $^{90}{\rm Ge}$. In fact, the resonance at $\epsilon_{\rm kin} \sim 0.03-0.07$ MeV corresponds to the peak
at $E=3.35$ MeV of the quadrupole strength function for the $L^{\pi}=2^{+}$ excited states in $^{90}{\rm Ge}$ (Fig.~\ref{strength_function_90Ge}(b)). 
Many of the other resonance structures also correspond
to the peaks in the quadrupole strength function, e.g., the small peak at $\epsilon_{\rm kin} \sim 3.4$ MeV corresponds to a peak at $E \sim 6.7$ MeV.

\begin{figure}[H]
\centering
\begin{minipage}{0.49\columnwidth}
\includegraphics[width=\columnwidth]
{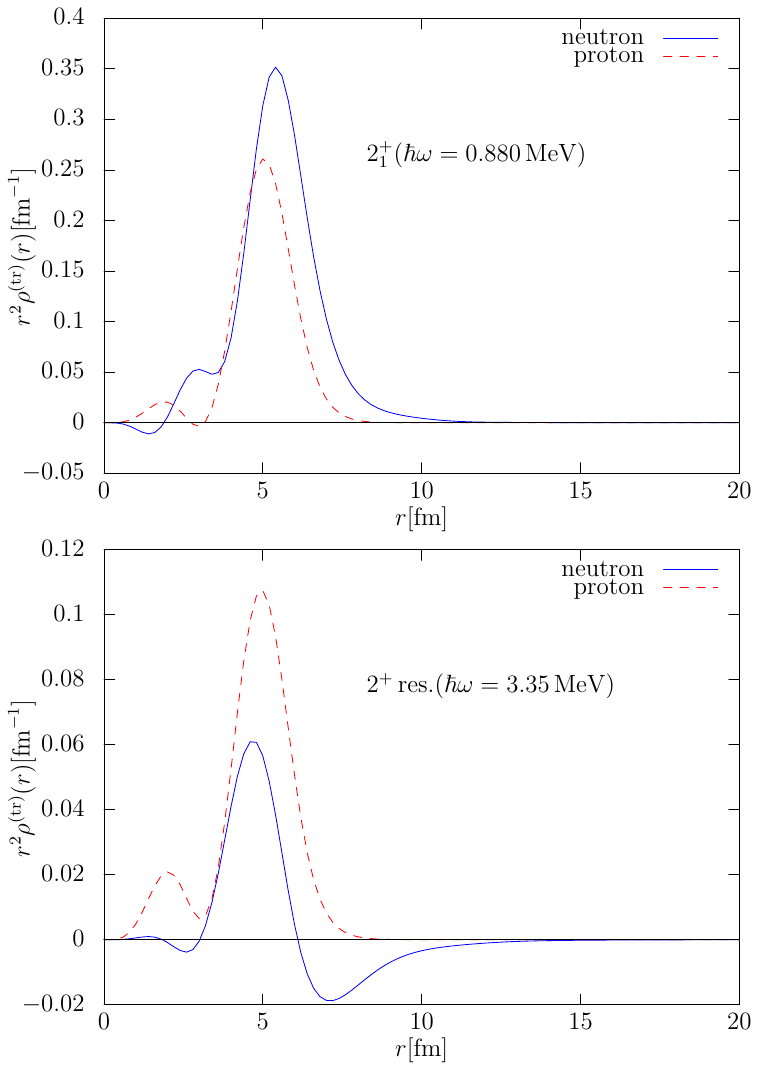}
\end{minipage} 
\caption{(a) The transition densities of neutrons and protons for the quadrupole vibrational state $2^{+}_{1}$ ($\hbar \omega=0.880$ MeV) of $^{90}{\rm Ge}$. (b) Same as (a) but for the pygmy
quadrupole resonance  (PQR, $\hbar \omega=3.35$ MeV) located just above the 
one-neutron separation energy.}
\label{transition_density_90Ge}
\end{figure}

The resonance $\epsilon_{\rm kin} \sim 0.03-0.07$ MeV
 is a collective state as seen in distribution of the forward amplitudes shown
 in Table \ref{amp_XY_PQR_90Ge}. But it has  
 a characteristic feature, different from the $2_1^{+}$ state of 
isoscalar surface vibration. 
A peak $E \approx 3.4$ MeV in the quadrupole strength function (Fig.~\ref{strength_function_90Ge}(b)), which corresponds 
 to  the resonance at $\epsilon_{\rm kin} \sim 0.03-0.07$ MeV,  exhibits isovector strength larger than isoscalar strength in contrast to other peaks. 
Figure \ref{transition_density_90Ge}(b) shows the  transition density of this state. It is seen that
neutrons oscillate with opposite phases in the inner and outer regions of the surface
 $r \sim 1.2 \times 90^{1/3}=5.4$ fm, and that neutrons and protons oscillate with opposite phase in the outer region
while they move coherently inside the surface. 
This behavior is the same as pygmy quadrupole resonance (PQR) predicted in Ref.~\cite{Tsoneva2011}, considered
as an analogy of the pygmy dipole resonance. It is contrasting to the isoscalar surface vibration, i.e., 
the $2_1^{+}$ state whose transition density is shown in Fig.~\ref{transition_density_90Ge}(a).
In the following we refer the resonance at $E \approx 3.4$ MeV 
($\epsilon_{\rm kin} \sim 0.03-0.07$ MeV) as a PQR. 

\begin{figure}[H]
\centering
\begin{minipage}{0.49\columnwidth}
\includegraphics[width=\columnwidth]
{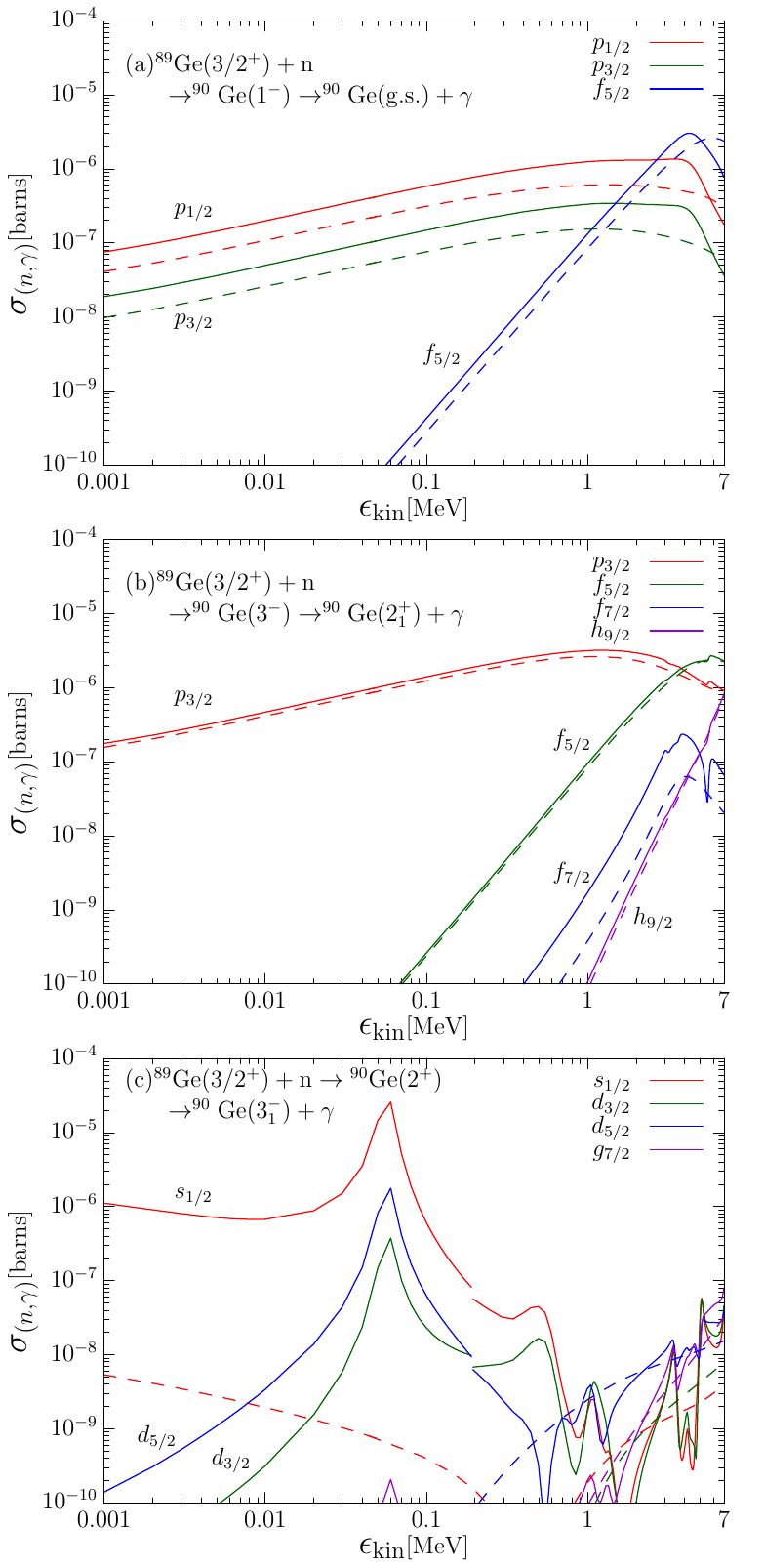}
\end{minipage}
\caption{(a) The calculated partial $(n,\gamma)$ cross sections of
the reaction $^{89}{\rm Ge}(3/2^{+})  +n \to ^{90}{\rm Ge}(1^{-}) \to  ^{90}{\rm Ge}({\rm g.s.}) + \gamma$ with $E1$ transition, plotted separately for different partial waves of the incident neutron. 
The solid curves represent the results of the present theory while the dashed curves do the results of the
semi-QRPA approximation, see text. 
The horizontal axis is kinetic energy $\epsilon_{\rm kin}$ of the incident neutron.
The smoothing parameter $\epsilon$ is $0.1$ MeV. 
(b) Same as (a) but for 
$^{89}{\rm Ge}(3/2^{+})  +n \to ^{90}{\rm Ge}(3^{-}) \to  ^{90}{\rm Ge}(2^{+}_{1}) + \gamma$.
(c) Same as (a) but for
 $^{89}{\rm Ge}(3/2^{+})  +n \to ^{90}{\rm Ge}(2^{+}) \to  ^{90}{\rm Ge}(3^{-}_{1}) + \gamma$.
 Small smoothing parameter $\epsilon=0.005$ MeV is used in low-energy region $\epsilon_{\rm kin} < 0.2$ MeV.}
\label{n_cap_partial_cross_section_90Ge_3/2+}
\end{figure}

The large impact of the PQR on the cross section has two origins. Because of the collectivity
the PQR state couples to (consists of) a large number of two-quasiparticle configurations, including
those 
in which one of the quasineutron is $2d_{3/2}$ (the configuration of
$^{89}{\rm Ge}(3/2^{+})$) and the other is scattering quasineutron in $s, d_{3/2}, d_{5/2}$ and $g_{7/2}$-waves.
A consequence is seen in 
Fig. \ref{n_cap_partial_cross_section_90Ge_3/2+}(c) showing
the cross sections decomposed with respect to the partial waves of incoming neutrons 
for $^{89}{\rm Ge}(3/2^{+})  +n \to ^{90}{\rm Ge}(2^{+}) \to ^{90}{\rm Ge}(3^{-}_{1}) + \gamma$,
where the spin-parity $L^\pi=2^{+}$ of the excited states of  $^{90}{\rm Ge}$ is specified. 
The enhanced cross section at the resonance around  $\epsilon_{\rm kin} \sim 0.03-0.07$ MeV are seen in
multiple partial waves $s_{1/2}, d_{3/2}$ and $d_{5/2}$, rather than in single partial wave. 
In particular, the  configuration including $s$-wave 
 has a significant impact on the cross section since the $s$-wave becomes dominant at low energy. 

Another important aspect of  PQR is that the PQR includes 
proton two-quasiparticle configurations as well as neutron configurations as shown in Table \ref{amp_XY_PQR_90Ge}. 
Therefore, the $E1$ operator acting
on the proton component contribute to the  transition between the PQR and the $3^{-}_{1}$ state:
two-quasiproton configurations participate to rings in the middle part of the diagram Fig.~\ref{diagram_90Ge}(b). In addition,
the isovector character of the PQR is preferable for enhancing the $E1$ transition as the $E1$ operator $\propto 
(N/A)\sum_{proton}(rY_{1m})-(Z/A)\sum_{neutron}(rY_{1m})$  is of isovector type.

\begin{table}[H]
\centering
\caption{Same as Table \ref{amp_XY_1st2+_90Ge} but for the PQR of $^{90}{\rm Ge}$. 
Two-quasiparticle configurations with large amplitude $|X_{ij}|>0.1$ are listed.}
\label{amp_XY_PQR_90Ge}
\begin{tabular}{crr p{5mm} crr} \\ \hline \hline 
neutron config. & \multicolumn{1}{c}{$X_{ij}$} & \multicolumn{1}{c}{$Y_{ij}$} && proton config. & \multicolumn{1}{c}{$X_{ij}$} & \multicolumn{1}{c}{$Y_{ij}$} \\ \hline
$[2d_{5/2}][2d_{3/2}]$ & 0.49 & -0.025 && $[2p_{3/2}]^{2}$ & -0.54 & -0.020 \\
$[1g_{7/2}][2d_{3/2}]$ & 0.38 & 0.038 && $[1f_{5/2}]^{2}$ & -0.27 & -0.011 \\
$[1g_{7/2}][2d_{5/2}]$ & -0.18 & -0.018 && $[1f_{5/2}][2p_{3/2}]$ & -0.24 & 0.002 \\
$[1g_{7/2}]^{2}$ & 0.15 & -0.028 && $[1f_{5/2}][2p_{1/2}]$ & 0.17 & 0.004 \\
$[2d_{5/2}][3s_{1/2}]$ & 0.13 & 0.034 && $[2p_{3/2}][2p_{1/2}]$ & -0.15 & 0.020 \\
\hline \hline
\end{tabular}
\end{table}

Figure \ref{n_cap_partial_cross_section_90Ge_3/2+}(a) shows the cross section for
 $^{89}{\rm Ge}(3/2^{+})  +n \to ^{90}{\rm Ge}(1^{-}) \to ^{90}{\rm Ge}({\rm g.s.}) + \gamma$ decaying
 to the ground state of $^{90}{\rm Ge}$ via $1^{-}$ excited states of the same nucleus. 
 There is no sharp resonances in this reaction channel, reflecting the dipole strength function (Fig.~\ref{strength_function_90Ge})
 which shows only a broad continuum bump (the soft dipole excitation) at low excitation energy. Main components
 of the soft dipole excitation in the present example are two-quasineutron excitations
 consisting of bound $2d_{3/2}$ and $3s_{1/2}$ quasiparticles and unbound (scattering)
 $p$- and $f_{5/2}$-waves. The $p$-waves dominate the capture
 cross section at low kinetic energy due to the centrifugal barrier. Note that the QRPA correlation enhances the cross section by
a factor of about two, indicating that the soft dipole excitation is not a pure single-particle transition \cite{Matsuo2005}.

 For the capture reaction with $E1$ transition to the $2_1^{+}$ state, the states with 
 total angular momentum $L^{\pi}=1^{-},2^{-},3^{-}$ are relevant. Similarly to the
 transition to the ground state, there is no sharp resonances. An example is shown
in Fig.~\ref{n_cap_partial_cross_section_90Ge_3/2+}(b)  for the reaction channel
 $^{89}{\rm Ge}(3/2^{+})  +n \to ^{90}{\rm Ge}(3^{-}) \to ^{90}{\rm Ge}(2_1^{+}) + \gamma$. 

\begin{figure}[H]
\centering
\begin{minipage}{0.49\columnwidth}
\includegraphics[width=\columnwidth]
{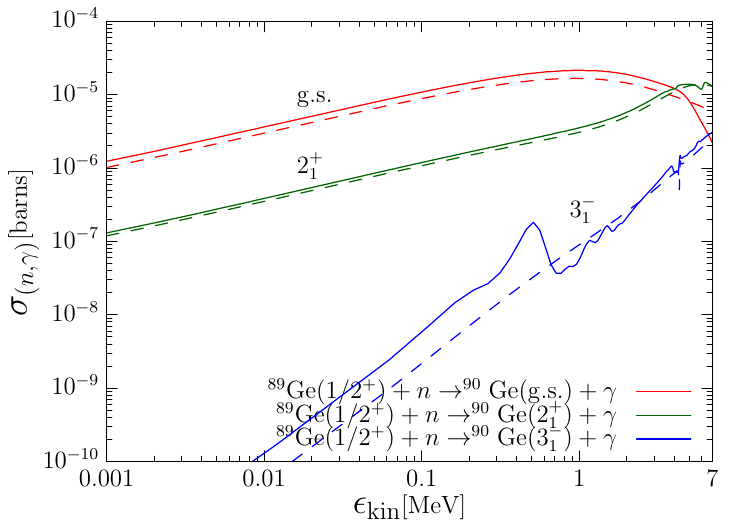}
\end{minipage} 
\caption{Same as Fig.~\ref{n_cap_total_cross_section_90Ge_3/2+} but for $^{89}{\rm Ge}(1/2^{+}) + n \to ^{90}{\rm Ge} + \gamma$.
The smoothing parameter $\epsilon$ is $0.1$ MeV.}
 \label{n_cap_total_cross_section_90Ge_1/2+}
\end{figure}

We shortly discuss the case where the configuration of the target nucleus $^{89}{\rm Ge}$ is the one-quasineutron
state $3s_{1/2}$ with spin-parity $1/2^{+}$.
Figure \ref{n_cap_total_cross_section_90Ge_1/2+} shows $E1$ $(n,\gamma)$ cross section 
$^{89}{\rm Ge}(1/2^{+}) + n \to ^{90}{\rm Ge}({\rm g.s.}, 2^{+}_{1}, 3^{-}_{1}) + \gamma$.
In this case, we only see a tiny resonance behavior: the small peak at $\epsilon_{\rm kin}=0.5$ MeV in the cross
section for the transition to the  $3^{-}_{1}$ state is caused by the PQR in the $2^{+}$ states. 
In this case, the $s$-wave capture for the path $^{89}{\rm Ge}(1/2^{+}) +n \to ^{90}{\rm Ge}(2^{+}) \to  ^{90}{\rm Ge}(3^{-}_{1}) + \gamma$ is 
prohibited due to the angular momentum conservation. Only the $d$-waves can form $2^{+}$ states, making the cross section very small.

\subsection{$^{91}{\rm Zn}(n,\gamma)^{92}{\rm Zn}$}

The low-lying quasiparticle states in $^{92}{\rm Zn}$ obtained with SkM$^{*}$ are listed in Table \ref{spe_92Zn}.  Since only the state $2d_{3/2}$ is a bound quasineutron state, 
we assume that
the ground state of $^{91}{\rm Zn}$ has the one-quasineutron  configuration
$2d_{3/2}$ with total spin-parity $3/2^{+}$. We then describe $(n,\gamma)$ reaction
in which a neutron is impinging on the ground state $^{91}{\rm Zn}(3/2^{+})$.
Note that a bound Hartree-Fock orbit $3s_{1/2}$ located just below the Fermi energy becomes
 a resonant quasiparticle state, called quasiparticle resonance \cite{Dobaczewski1996,Bennaceur1999,Balgac2000}, which is
 embedded in continuum just above (by $\sim 0.2$ MeV) the threshold energy $E = |\lambda_{n}| = 1.43$ MeV.
 
\begin{figure}[H]
\centering
\begin{minipage}{0.49\columnwidth}
\includegraphics[width=\columnwidth]
{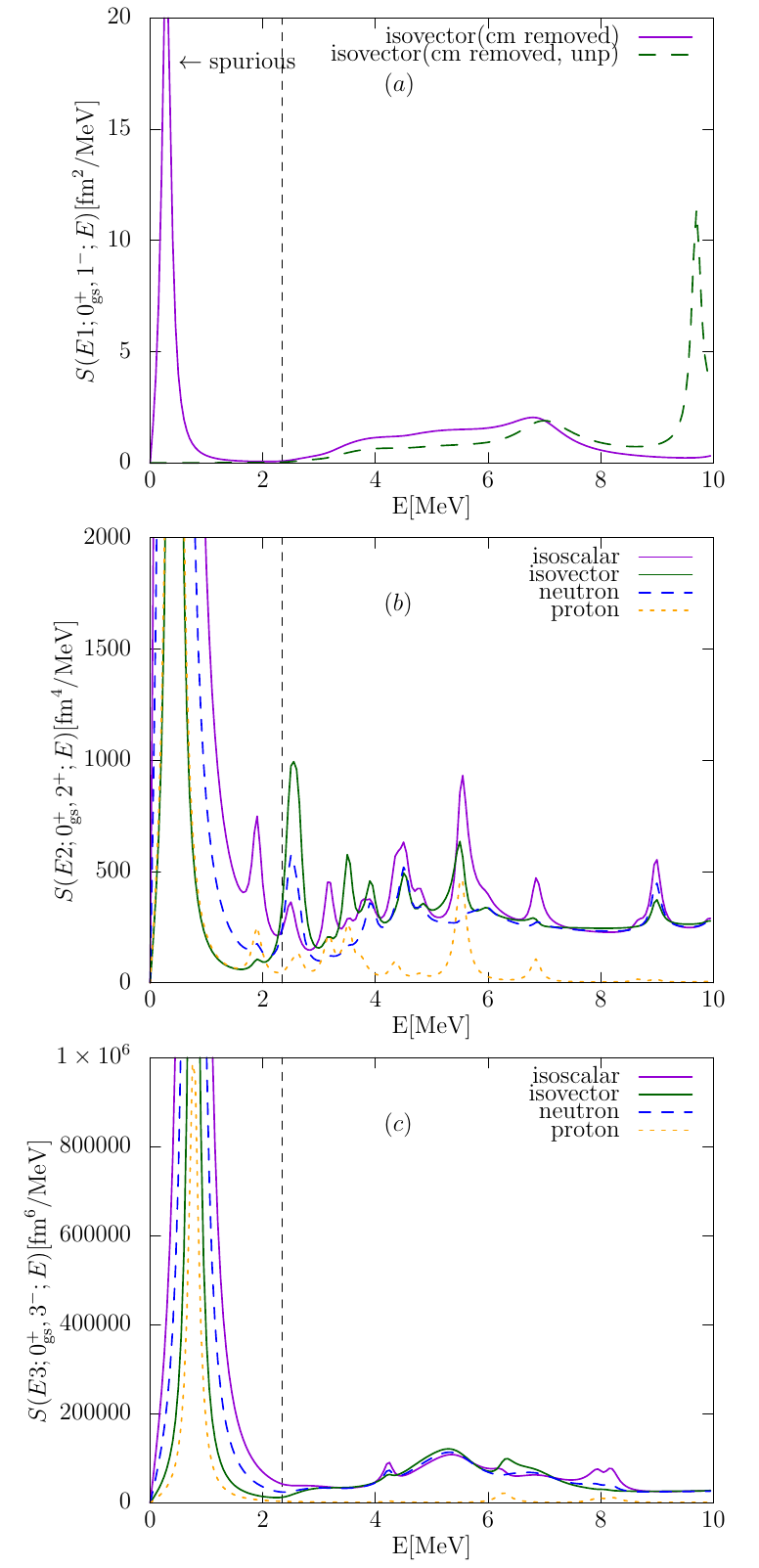}
\end{minipage} 
\caption{(a) The strength function for $L^{\pi}=1^{-}$ excited states in $^{92}{\rm Zn}$, calculated with $\epsilon=0.1$ MeV.
The horizontal axis is the excitation energy of the $1^{-}$ states.
The vertical dashed line is one neutron separation energy $S_{1n}(2d_{3/2})=2.34$ MeV. 
(b) and (c) Same as (a) but for $L^{\pi}=2^{+}$ and $3^{-}$ states.}
\label{strength_function_92Zn}
\end{figure}

Figure \ref{strength_function_92Zn} shows the dipole, quadrupole and octupole strength functions for multipole
operators, calculated for $1^{-},2^{+},3^{-}$ excited sates in $^{92}{\rm Zn}$ in the excitation energy range
 $E=0-10$ MeV. Significant low-energy peaks are seen at $E_{2^{+}_{1}}=0.380$ and $E_{3^{-}_{1}}=0.950$ MeV in the quadrupole and octupole strength functions, respectively. These are collective states as 
 evident from the forward and backward amplitudes $X_{ij},Y_{ij}$, listed in Tables \ref{amp_XY_1st2+_92Zn} and \ref{amp_XY_1st3-_92Zn}.

\begin{table}[H]
\centering
\caption{Single-quasiparticle energies of the adopted Skyrme energy density functional SkM$^{*}$ for $^{92}{\rm Zn}$, obtained with the
box boundary condition.
Quasiparticle states corresponding to bound Hartree-Fock orbits around the Fermi energy ($\lambda_{n}= -1.43, \lambda_{p}=-18.40$ MeV) are listed.
Quasineutron states $3s_{1/2}, 1g_{7/2}$ and $2d_{5/2}$, whose energies are larger
than the threshold $|\lambda_{n}|= 1.43$ MeV, are quasiparticle resonances.}
\label{spe_92Zn}
\begin{tabular}{cc p{5mm} cc} \\ \hline \hline
neutrons & $E_i$ (MeV) && protons & $E_i$ (MeV) \\ \hline
$2d_{5/2}$ & 3.03 && $1g_{9/2}$ & 5.79 \\
$1g_{7/2}$ & 1.77 && $1f_{7/2}$ & 4.53 \\
$3s_{1/2}$ & 1.67 && $2p_{1/2}$ & 2.65 \\
$2d_{3/2}$ & 0.91 &&  $1f_{5/2}$ & 2.35 \\
& && $2p_{3/2}$ & 1.34 \\
\hline \hline
\end{tabular}
\end{table}

\begin{table}[H]
\centering
\caption{Same as Table \ref{amp_XY_1st2+_90Ge} but for the $2^{+}_{1}$ of $^{92}{\rm Zn}$. 
Two-quasiparticle configurations with the largest 10 amplitudes are listed.
See also the caption in Table \ref{amp_XY_1st3-_90Ge}.}
\label{amp_XY_1st2+_92Zn}
\begin{tabular}{crr p{5mm} crr} \\ \hline \hline 
neutron config. & \multicolumn{1}{c}{$X_{ij}$} & \multicolumn{1}{c}{$Y_{ij}$} && proton config. & \multicolumn{1}{c}{$X_{ij}$} & \multicolumn{1}{c}{$Y_{ij}$} \\ \hline
$[1g_{7/2}][2d_{3/2}]$ & 0.76 & -0.53 && $[2p_{3/2}]^{2}$ & -0.44 & 0.31 \\
$[2d_{3/2}]^{2}$ & 0.47 & -0.27 && $[2p_{3/2}][2p_{1/2}]$ & -0.35 & -0.27 \\
$[1g_{7/2}]^{2}$ & 0.38 & -0.27 && $[1f_{7/2}][2p_{3/2}]$ & -0.34 & 0.29 \\
$[c_{1}h_{11/2}]^{2}$ & -0.31 & 0.23 && $[1f_{5/2}][2p_{3/2}]$ & -0.24 & -0.18 \\
$[2d_{3/2}][3s_{1/2}]$ & 0.26 & 0.16 && $[1f_{5/2}][2p_{1/2}]$ & 0.23 & -0.18 \\
$[c_{2}i_{13/2}][1g_{9/2}]$ & -0.15 & 0.14 && $[1f_{5/2}]^{2}$ & -0.20 & 0.15 \\
$[1f_{7/2}][c_{1}h_{11/2}]$ & -0.15 & 0.14 && $[1h_{11/2}][1f_{7/2}]$ & -0.17 & 0.17 \\
$[1g_{7/2}][2d_{5/2}]$ & -0.12 & -0.099 && $[1d_{5/2}][1g_{9/2}]$ & -0.12 & 0.11 \\
$[2d_{5/2}][2d_{3/2}]$ & 0.11& 0.078 && $[1f_{7/2}][1f_{5/2}]$ & 0.11 & 0.095 \\
$[c_{3}i_{13/2}][1g_{9/2}]$ & -0.11 & 0.11 && & & \\
\hline \hline
\end{tabular}
\end{table}

\begin{table}[H]
\centering
\caption{Same as Table \ref{amp_XY_1st2+_90Ge} but for the $3^{-}_{1}$ of $^{92}{\rm Zn}$. 
Two-quasiparticle configurations with the largest 10 amplitudes are listed.
See also the caption in Table \ref{amp_XY_1st3-_90Ge}.}
\label{amp_XY_1st3-_92Zn}
\begin{tabular}{crr p{5mm} crr} \\ \hline \hline 
neutron config. & \multicolumn{1}{c}{$X_{ij}$} & \multicolumn{1}{c}{$Y_{ij}$} && proton config. & \multicolumn{1}{c}{$X_{ij}$} & \multicolumn{1}{c}{$Y_{ij}$} \\ \hline
$[c_{1}h_{11/2}][2d_{5/2}]$ & 0.73 & -0.51 && $[1g_{9/2}][2p_{3/2}]$ & 0.53 & -0.39 \\
$[1g_{9/2}][c_{1}h_{11/2}]$ & 0.45 & -0.36 && $[1g_{9/2}][1f_{7/2}]$ & 0.42 & -0.35 \\
$[c_{1}d_{5/2}][c_{1}h_{11/2}]$ & -0.30 & 0.21 && $[c_{1}i_{13/2}][1f_{7/2}]$ & -0.19 & 0.18 \\
$[c_{1}h_{11/2}][1g_{7/2}]$ & 0.24 & 0.14 && $[1f_{5/2}][2s_{1/2}]$ & -0.19 & -0.16 \\
$[2p_{1/2}][1g_{7/2}]$ & 0.23 & -0.20 && $[1f_{7/2}][2d_{5/2}]$ & -0.17 & 0.15 \\
$[c_{2}f_{7/2}][3s_{1/2}]$ & 0.19 & -0.14 && $[1d_{3/2}][2p_{3/2}]$ & -0.16 & -0.13 \\
$[1f_{5/2}][1g_{7/2}]$ & 0.19 & -0.15 && $[1f_{5/2}][1d_{3/2}]$ & -0.16 & 0.13 \\
$[c_{1}f_{7/2}][3s_{1/2}]$ & 0.18 & -0.12 && $[1d_{5/2}][1h_{11/2}]$ & -0.15 & 0.14 \\
$[c_{3}h_{9/2}][2d_{3/2}]$ & 0.16 & -0.14 && $[2d_{3/2}][2p_{3/2}]$ & -0.15 & -0.13 \\
$[c_{2}f_{7/2}][2d_{5/2}]$ & -0.16 & 0.12 && $[2d_{5/2}][2p_{3/2}]$ & -0.14 & 0.12 \\
\hline \hline
\end{tabular}
\end{table}

\begin{figure}[H]
\centering
\begin{minipage}{0.49\columnwidth}
\includegraphics[width=\columnwidth]
{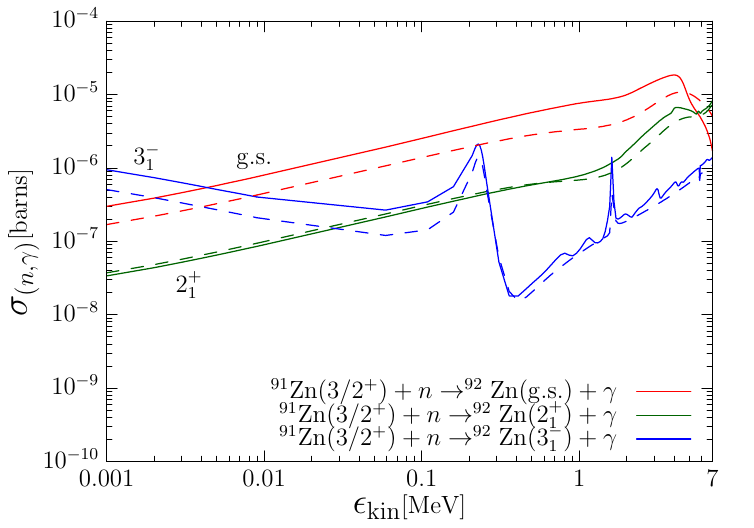}
\end{minipage} 
\caption{
Same as Fig.~\ref{n_cap_total_cross_section_90Ge_3/2+} but for $^{91}{\rm Zn}(3/2^{+}) +n \to ^{92}{\rm Zn} + \gamma$.
The smoothing parameter $\epsilon$ is $0.1$ MeV.}
\label{n_cap_total_cross_section_92Zn}
\end{figure}

Figure \ref{n_cap_total_cross_section_92Zn} shows  calculated capture cross section for the $(n,\gamma)$ reaction
$^{91}{\rm Zn}(3/2^{+}) + n \to ^{92}{\rm Zn}({\rm g.s.}, 2^{+}_{1}, 3^{-}_{1}) + \gamma$
in which the final states of the $E1$ transition 
are the low-lying $2^{+}_{1}$ and $3^{-}_{1}$
states as well as the ground state of $^{92}{\rm Zn}$.  The cross section for the 
transition to the  $3^{-}_{1}$ state is noticeable as it exhibits an
 apparent resonant behavior at $\epsilon_{\rm kin} \sim 0.2$ MeV. 
 This resonance arises from  the excited $2^{+}$
state of  $^{92}{\rm Zn}$ in the continuum, corresponding to
a peak at $E \sim 2.6$ MeV in the quadrupole strength function
 (Fig.~\ref{strength_function_92Zn}(b)).
 We notice further that the cross section increases with decreasing the neutron 
energy below the resonance energy $\epsilon_{\rm kin} \lesim 0.1$ MeV, and becomes larger than the cross section for the 
transition to the ground state. 
There exist also many other resonant behaviors at higher energy
$\epsilon_{\rm kin}=1-7$ MeV. These resonances correspond to the peaks in the
range of 
$2-9$ MeV in the quadrupole strength function and non-correlated $3^{+}$, correlated $4^{+}$ peaks.

\begin{figure}[H]
\centering
\begin{minipage}{0.49\columnwidth}
\includegraphics[width=\columnwidth]
{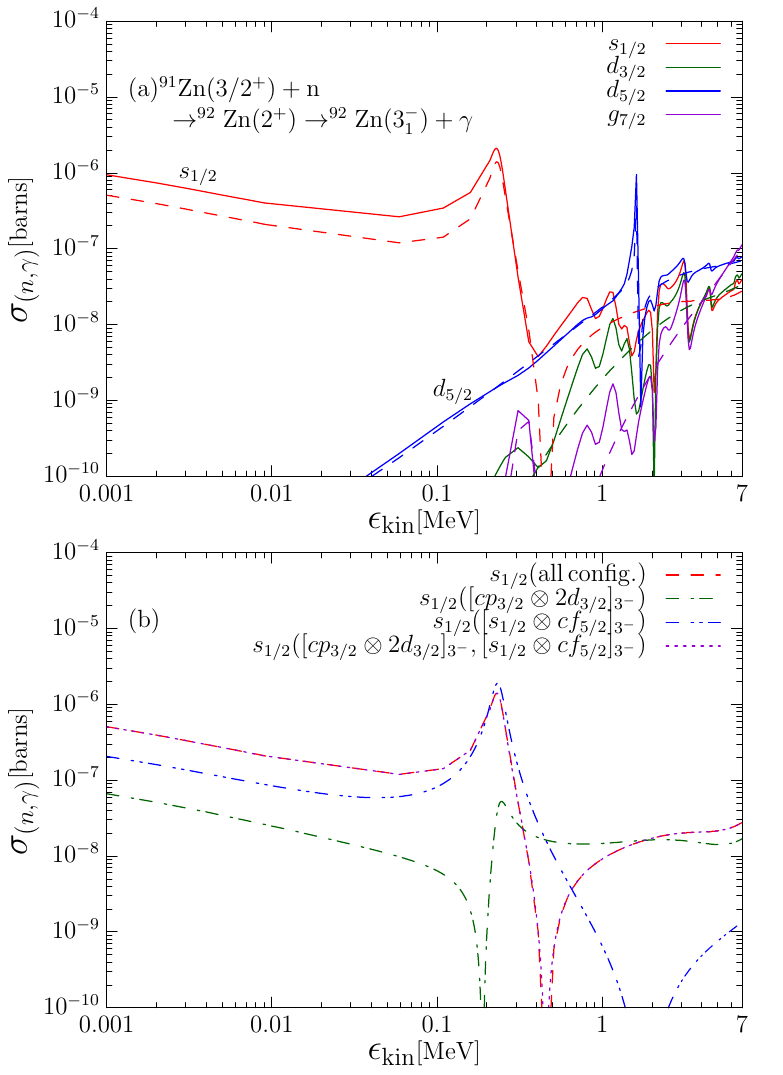}
\end{minipage} 
\caption{(a) The calculated partial $(n,\gamma)$ cross sections for the specific channel $^{91}{\rm Zn}(3/2^{+})  +n \to ^{92}{\rm Zn}(2^{+}) \to ^{92}{\rm Zn}(3^{-}_{1}) + \gamma$ with $E1$ transition, plotted separately for different partial waves of the incident neutron.
The solid curves represent the results of the present theory while the dashed curves do the results of the
semi-QRPA approximation, see text. 
(b) The calculated semi-QRPA partial $(n,\gamma)$ cross sections where scattering neutron is limited to $s$-wave 
and two-quasineutron configurations are limited to  $[cp_{3/2} \otimes 2d_{3/2}]_{3^{-}}$, $[s_{1/2} \otimes cf_{5/2}]_{3^{-}}$ or both.
The horizontal axis is kinetic energy $\epsilon_{\rm kin}$ of the incident neutron.
The smoothing parameter $\epsilon$ is $0.1$ MeV.}
\label{n_cap_partial_cross_section_92Zn}
\end{figure}

We note that the resonance at $\epsilon_{\rm kin} \sim 0.2$ MeV has different structure from the
one we have discussed for $^{90}{\rm Ge}$, in which case the QRPA
correlation played the key role to produce the resonant state, i.e., the PQR.
This is seen from a comparison with the
semi-QRPA calculation (dotted curve). 
There exists essentially the same resonance behavior at $\epsilon_{\rm kin} \sim 0.2$ MeV in
the semi-QRPA calculation although the QRPA correlation gives only a weak enhancement
of the cross section around this resonance (and at lower energies) by a factor
of about two. Figure \ref{n_cap_partial_cross_section_92Zn}(a) shows decomposition
of the cross section with respect to the partial waves of the incident neutron.
The resonance structure at $\epsilon_{\rm kin} \sim 0.2$ MeV appears
only in the $s$-wave neutron while it is not seen for the $d$-waves, in which 
a peak structure would have appeared if the origin of the resonance is the QRPA correlation.
The above observations suggest that
the resonance at $\epsilon_{\rm kin}\sim 0.2$ MeV is the one with single-particle character 
associated with the scattering $s$-wave neutron state. Indeed the peak of the
dotted line exactly corresponds to the quasineutron resonance 
with $E \approx 2.58$ MeV, originating from the
$3s_{1/2}$ orbit.

\begin{figure}[H]
\centering
\begin{minipage}{0.50\columnwidth}
\includegraphics[width=\columnwidth]
{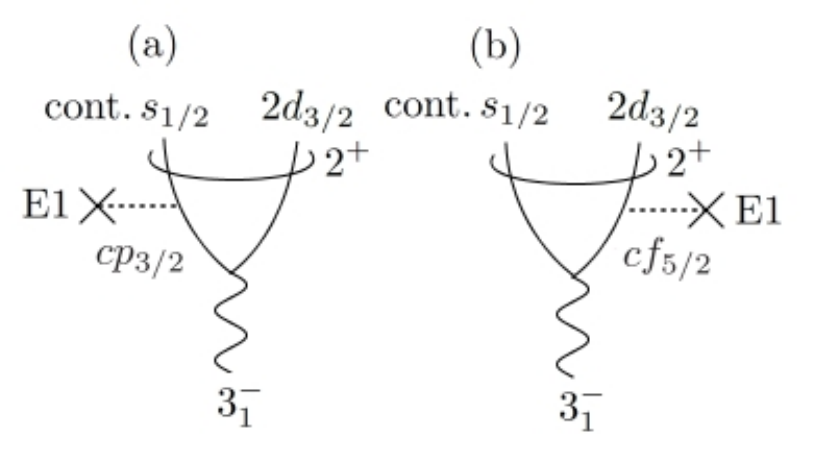}
\end{minipage} 
\caption{Diagrams in the semi-QRPA representing  $s$-wave  capture  $^{91}{\rm Zn}(3/2^{+}) +n_{s_{1/2}} \to ^{92}{\rm Zn}(2^{+}) \to ^{92}{\rm Zn}(3^{-}_{1}) + \gamma$ with $E1$ transition.}
\label{diagram_92Zn}
\end{figure}

Roles of the $3s_{1/2}$ quasineutron resonance may be explained 
in terms of the  diagrams shown in Fig.~\ref{diagram_92Zn}. 
The initial state of the reaction corresponds to the top part of the diagram, and it
 consists of the scattering $s$-wave quasineutron
and $2d_{3/2}$ quasineutron state.
This configuration has $E1$ matrix elements to the the $3_{1}^{-}$ state (represented by
the wavy line in the bottom part of the diagram).
There are two contributions, in which
the $E1$ operator acts on the $s$-wave neutron and  the $2d_{3/2}$ quasineutron state,
corresponding to the  diagrams (a) and (b), respectively.  The $E1$ matrix
elements are expressed as 
\begin{align}
\sum_{c} \langle s_{1/2} | E1 |c p_{3/2}\rangle X_{c p_{3/2},d_{3/2}}^{3_1^{-}} & \quad {\rm for\ (a)}, \notag \\
\sum_c \langle 2d_{3/2} | E1 |cf_{5/2}\rangle X_{s_{1/2},cf_{5/2}}^{3_1^{-}} & \quad {\rm for\ (b)}, 
\label{E1matrixel}
\end{align}
where $X_{ij}^{3_1^{-}} $ is the forward amplitude of two-quasiparticle configuration
$ij$ in the $3_{1}^{-}$ state. Note that the two-quasineutron configurations
$[cp_{3/2 }\otimes2d_{3/2}]_{3^{-}}$  contribute to the diagram (a) while
$[s_{1/2} \otimes cf_{5/2}]_{3^{-}}$ for (b).  
 Here $s_{1/2}$ is the scattering $s$-wave quasineutron, and $cp_{3/2}$ and
$cf_{5/2}$ denote the unbound quasineutron states (labeled with "$c$") with quantum numbers
$p_{3/2}$ and $f_{5/2}$.
The above two-quasineutron components
are present in the  $3_{1}^{-}$ state due to a large collectivity of this state
although the associated amplitudes $X_{ij}$ are small ($|X_{ij}| <0.1$). 

Figure \ref{n_cap_partial_cross_section_92Zn}(b) shows cross sections in which the contributions from
diagrams (a) and (b) are calculated separately. It is seen that
both components contribute to the resonance structure, but in different
ways.   The diagram (b) forms a peak structure at the resonance energy
while  (a) contributes to the cross section with different signs 
depending on whether the energy of incident neutron is above or below
the peak energy.

To understand this we note that $[s_{1/2} \otimes cf_{5/2}]_{3^{-}}$  components in the low-lying $3_{1}^{-}$ state
contribute to (b) while  $[cp_{3/2} \otimes 2d_{3/2}]_{3^{-}}$ components are relevant to (a).
For (b), the forward amplitudes $X_{ij}$ of the
$[s_{1/2}\otimes cf_{5/2}]_{3^{-}}$  components  become large in Eq.~(\ref{E1matrixel})
in the case when the scattering $s_{1/2}$ quasineutron is on the resonance.
(In this case two-quasineutron configuration $[s_{1/2}\otimes cf_{5/2}]_{3^{-}}$ becomes
that of a particle-hole character, a preferable feature for the vibrational $3^{-}_{1}$ state.
The $3s_{1/2}$ quasineutron resonance has a hole
character while unbound  $cf_{5/2}$ has a particle character.)
Hence the diagram  (b)  forms a peak at the resonance. 
For (a),  relevant  $E1$ matrix element is
\begin{align}
\langle s_{1/2} | E1 |cp_{3/2}\rangle &= \frac{1}{2}\int dx \phi^{\dag}_{s_{1/2}}(x) {\cal M}_{E1}(x) \phi_{cp_{3/2}}(x)  \notag\\
&=\frac{1}{2}\int dx f_{E1}(x)\left(\varphi^{*}_{s_{1/2},1}(x)\varphi_{cp_{3/2},1}(x)
- \phi^{*}_{s_{1/2},2}(x)\phi_{cp_{3/2},2}(x)\right),
\end{align}
for which the the upper component $\varphi^{*}_{s_{1/2},1}(x)$ of the $s_{1/2}$ quasineutron (scattering neutron)
gives dominant contribution as the unbound quasineutron state $cp_{3/2}$ has dominant upper
component $|\varphi_{cp_{3/2},1}(x)| \gg |\varphi_{cp_{3/2},2}(x)|$.
The upper component  $\varphi_{s_{1/2},1}(x)$ of the $s_{1/2}$
quasiparticle is a scattering wave, and its amplitude
changes sign depending on below or above the resonance energy, and hence contribution of
the diagram (a) also changes sign accordingly.

The components of (a) and (b) contribute coherently (destructively) to the cross section below (above) 
the resonance peak,
and this gives a large enhancement of the capture cross section at the lower tail
of the resonance.
We note that the diagram (b) is not usually considered in the single-particle neutron capture model because it only takes into account an incident neutron directly decaying to a bound single-particle orbit.
  
Summarizing,  the resonant behavior at $\epsilon_{\rm kin} \sim 0.2$ MeV and the enhancement
of the cross section at very low energy originate from a combined effect of the
quasiparticle resonance in the $s$-wave scattering neutron and the collectivity
of the low-lying $3_1^{-}$ state. 

\section{conclusions} \label{conclusions}
The cQRPA enables one to describe collective excitations which decay through
 emission of a nucleon. 
Taking this advantage of the cQRPA, we have  formulated a prototype theory previously \cite{Matsuo2015} with which one can
calculate the cross section of radiative neutron capture reaction on neutron-rich nuclei, relevant to $r$-process nucleosynthesis. In this prototype formulation, however,  
we have restricted ourselves to the case where 
the final state of $\gamma$ transition is  the ground state of an even-even nucleus.  In the present study, we have
extended the formulation of Ref.~\cite{Matsuo2015} by incorporating the method of Ref.~\cite{Saito2023}
that allows to describe the  $\gamma$ transitions 
populating low-lying excited states such as surface vibrational $2_1^+$ and $3_1^-$ states. In the present
framework, there appear resonances of collective and non-collective characters at low excitation energy
near the threshold of neutron separation. Contributions of the resonances to  the $(n,\gamma)$ cross section
are taken into account in a consistent way together with direct transitions from scattering neutron state.

We have demonstrated the new features of the theory by presenting numerical calculations performed for reactions on
neutron-rich open-shell nuclei in the region with $A\sim 90$ ( $N>50$, $Z>28$ ) with $S_{1n} \sim 2$ MeV:
$^{89}{\rm Ge}(n,\gamma)^{90}{\rm Ge}$ and $^{91}{\rm Zn}(n,\gamma)^{92}{\rm Zn}$, in which  
$E1$ $\gamma$ transition to the ground state or low-lying excited states $2^{+}_{1},3^{-}_{1}$ are evaluated.
The example of germanium is a case where a collective state having the character of the 
pygmy quadrupole resonance (PQR) appears in $^{90}{\rm Ge}$ just above the neutron separation energy. The resonance brings about 
a significant peak in the capture cross section as well as enhancement at very low energy. Here the enhanced cross
section occurs for the transition to the low-lying $3^{-}_{1}$ state of the octupole vibration.
Collectivities in the 
PQR and the octuple $3_1^{-}$ state play key roles. 

The case of zirconium provides an example in which another
type of resonance shows up. The neutron $3s_{1/2}$ orbit, which is an occupied bound orbit in the Hartree-Fock approximation,  
becomes with the neutron pairing a resonant quasiparticle state coupled to scattering neutron in the $s$-wave. 
This quasiparticle resonance contributes also to the cross section of $^{91}{\rm Zn}(n,\gamma)^{92}{\rm Zn}$,
especially for the transition to the octupole $3^{-}_{1}$ state in $^{92}{\rm Zn}$. Note that the $E1$ transition 
from a $s$-wave is possible only to  $p$-wave orbits, but there is no bound $p$-orbits around the 
$N \sim 60$ Fermi energy. Thanks to the collectivity,  the octupole $3^{-}_{1}$ state contains a large number
of two-quasineutron configurations, including those with unbound $p$-wave orbits, through which the
$E1$ transition becomes possible.

The numerical examples point to new resonance mechanisms in the low-energy radiative neutron
capture on neutron-rich nuclei. However,  quantitative aspects should be taken with reservation 
as the results depend on the Skyrme and pairing parameters. Dependence on these parameters will
be discussed in a forthcoming paper.  
In the present study, we have evaluated the $E1$ transitions. Inclusion of $M1$ component, another dominant transition, 
is straightforward.   For systematic description of the $(n,\gamma)$ reaction,
extension to other types of target nuclei, e.g., the proton-odd
and even-even nuclei, is necessary, and it is also in progress.

\section{Acknowledgments}
The numerical calculations were performed on the supercomputer HPE SGI8600 in the Japan Atomic Energy Agency.
This work was supported by the JSPS KAKENHI (Grant No.~20K03945 and No.~24K07014).

\appendix

\section{linear response equations}

The generalized density matrix is defined and evaluated as
\begin{align}
{\cal R}(x,y) = 
\begin{pmatrix}
 \langle \psi^{\dag}(y) \psi(x)\rangle &  \langle \psi(\tilde{y}) \psi(x)\rangle \\
 \langle \psi^{\dag}(y) \psi^{\dag}(\tilde{x})\rangle &  \langle\psi(\tilde{y})\psi^\dagger(\tilde{x})\rangle \\
\end{pmatrix}
=\sum_i \bar{\phi}_{\tilde{i}}(x)\bar{\phi}^\dagger_{\tilde{i}}(y).
\end{align}
Its fluctuation under the perturbing field $\hat{V}_{\rm scf}$ is then given as 
\begin{align}
\delta {\cal R}(x,y,\omega) & = \sum_i \delta \bar{\phi}_{\tilde{i}}(x,\omega)\bar{\phi}^\dagger_{\tilde{i}}(y) +\bar{\phi}_{\tilde{i}}(x)\delta\bar{\phi}^\dagger_{\tilde{i}}(y,-\omega) \notag \\ 
&= \iint dx^{'}dy^{'}\sum_i  {\cal G}_{0}(x,x^{'}, - E_{i} + \hbar \omega + i \epsilon){\cal V}_{\rm scf}(x^{'}, y^{'},\omega) 
\bar{\phi}_{\tilde{i}}(y^{'})\bar{\phi}^\dagger_{\tilde{i}}(y) \notag \\ 
&+\iint dx^{'}dy^{'}\sum_i  \bar{\phi}_{\tilde{i}}(x)\bar{\phi}^\dagger_{\tilde{i}}(x^{'}){\cal V}_{\rm scf}(x^{'}, y^{'},\omega) {\cal G}_{0}(y^{'},y, - E_{i} - \hbar \omega - i \epsilon),
\end{align}
where ${\cal G}_0(E)= (E - {\cal H}_{0})^{-1}$ is the quasiparticle Green's function defined with the $2 \times 2$ Hamiltonian ${\cal H}_{0}$
for the HFB equation, Eq.~(\ref{HFB_equation}).
Each component of the generalized density matrix fluctuation is accordingly given by
\begin{align}
\delta \rho_\alpha(x,y,\omega) = {\rm Tr} {\cal A}_{\alpha} \delta{\cal R}(x,y,\omega)
=\iint dx^{'}dy^{'} \sum_{\beta} R^{\alpha \beta}_{0}(x,y;y^{'},x^{'};\omega) V^{\rm scf}_{\beta}(x^{'},y^{'},\omega).
\end{align}
The function $R^{\alpha \beta}_{0}(x,y;y^{'},x^{'};\omega)$ is an unperturbed response function defined by 
\begin{align} \label{R_0_Green}
R^{\alpha \beta}_{0}(x,y;y^{'},x^{'};\omega) &= \sum_{i} \Big\{ \bar{\phi}^{\dag}_{\tilde{i}}(y)  {\cal A}_{\alpha} {\cal G}_{0}(x, x^{'}, - E_{i} + \hbar \omega + i \epsilon) {\cal A}_{\beta} \bar{\phi}_{\tilde{i}}(y^{'}) \notag \\
&+ \bar{\phi}^{\dag}_{\tilde{i}}(x^{'}) {\cal A}_{\beta} {\cal G}_{0}(y^{'}, y, - E_{i} - \hbar \omega - i \epsilon) {\cal A}_{\alpha} \bar{\phi}_{\tilde{i}}(x) \Big\}.
\end{align}

Furthermore, the response function is expressed as a contour integral form,
\begin{align} \label{R_0_continuum}
R^{\alpha \beta}_{0}(x,y;y^{'},x^{'};\omega) &= \frac{1}{2 \pi i} \int_{C} dE \Big\{ {\rm Tr} {\cal A}_{\alpha} {\cal G}_{0}(x, x^{'}, E + \hbar \omega + i \epsilon) {\cal A}_{\beta} {\cal G}_{0}(y^{'}, y,E) \notag \\
&+ {\rm Tr} {\cal A}_{\alpha} {\cal G}_{0}(x, x^{'}, E) {\cal A}_{\beta} {\cal G}_{0} (y^{'}, y, E- \hbar \omega - i \epsilon) \Big\}.
\end{align}
The Green's function satisfies proper asymptotic boundary condition for $|{\bf r}_{x}|,|{\bf r}_{x^{'}}| \to \infty$ so that describes unbound scattering waves \cite{Matsuo2001}. In the actual numerical calculation we use this expression in order to treat scattering states properly without
discretization of spectrum. 

The spectral representation of the unperturbed response function  $R^{\alpha \beta}_{0}(x,y;y^{'},x^{'};\omega)$ given by Eq.~(\ref{R_0_spectrum})
is obtained by using the spectral representation of the quasiparticle Green's function,
\begin{align}
{\cal G}_0(x,x^{'},E)=\sum_i \frac{\phi_i(x)\phi_i^\dagger(x^{'})}{E-E_i} + \frac{ \bar{\phi}_{\tilde{i}}(x)\bar{\phi}_{\tilde{i}}^\dagger(x^{'})}{E+E_i},
\end{align}
and $\bra{ij}\hat{V}\ket{0}=\iint dx dy \phi_i^\dagger(x){\cal V}(x,y)\bar{\phi}_{\tilde{j}}(y)$ for two-quasiparticle state
$\ket{ij}$.

In the numerical calculations presented in this paper, we assume that the density functional which produces the mean field
$\hat{U}$ and the induced field
$\delta \hat{U}_{\rm ind}$ is a functional of spin-independent local densities:
\begin{align}
\rho_\alpha(\vecr)=\sum_\sigma\langle \psi^{\dag}(x) \psi(x) \rangle, \sum_\sigma\langle \psi(\tilde{x}) \psi(x) \rangle,  \sum_\sigma\langle \psi^{\dag}(x) \psi^{\dag}(\tilde{x}) \rangle,
\end{align}
representing particle density and pair-add and pair-removal densities, respectively.
Then the matrix elements of the induced field $\delta \hat{U}_{\rm ind}$ read
\begin{align}
\delta U_\beta^{\rm ind}(x,y,\omega) = \sum_\gamma \frac{\del U_\beta[\rho]}{\del\rho_\gamma(\vecr)}\delta\rho_\gamma(\vecr,\omega) \delta(x-y),
\end{align}
and the fluctuations in the local densities follow
\begin{align}
\delta\rho_\alpha(\vecr,\omega) =&\sum_{\sigma} \int dx^{'} \sum_{\beta}
R^{\alpha \beta}_{0}(x,x;x^{'},x^{'};\omega) \sum_{\gamma} \frac{\del U_\beta[\rho]}{\del\rho_\gamma(\vecr')}\delta\rho_\gamma(\vecr',\omega) \notag \\ 
&+ \sum_{\sigma} \iint dx^{'}dy^{'} \sum_{\beta} R^{\alpha \beta}_{0}(x,x;y^{'},x^{'};\omega)F_\beta(x^{'},y^{'}).
\end{align}
Solving this equation, we obtain density fluctuations $\delta\rho_\alpha(\vecr,\omega)$ and also the induced field $\delta \hat{U}_{\rm ind}(\omega)$

\section{Equations with spherical symmetry}

We give here some equations where the system is spherical symmetric, and the partial wave representations are adopted
with  the polar coordinate system.

We express the quasiparticle wave function as 
\begin{align}
\phi(x) = Y_{ljm}(\hat{x})\phi_{lj}(r_x)/r_x,
\end{align}
and the quasiparticle Green's function as 
\begin{align}
{\cal G}_0(x,x^{'},E) = \sum_{ljm} Y_{ljm}(\hat{x}) \frac{1}{r_{x} r_{x^{'}}} {\cal G}_{0, lj}(r_{x}, r_{x^{'}}, E) Y^{*}_{ljm}(\hat{x}^{'}).
\end{align}
The response functions are expanded as 
\begin{align}
R_0^{\alpha\beta}(x,y; y^{'},x^{'}; \omega) = \sum_{ljm, l^{'}j^{'}m^{'}} Y_{l^{'}j^{'}m^{'}}(\hat{x}) Y^{*}_{ljm}(\hat{y}) \frac{1}{r_{x} r_{y} r_{y^{'}} r_{x^{'}}} R^{\alpha \beta}_{0, l^{'}j^{'}, lj}(r_{x}, r_{y}; r_{y^{'}}, r_{x^{'}}; \omega) Y_{ljm}(\hat{y}^{'}) Y^{*}_{l^{'}j^{'}m^{'}}(\hat{x}^{'}).
\end{align}
Here $Y_{ljm}$ is the spin spherical harmonics.

The density responses with multipolarity $LM$ is expressed as
 \begin{align}
\delta\rho_{\alpha LM}(x, y, \omega) = \sum_{ljm, l^{'}j^{'}m^{'}} Y_{l^{'}j^{'}m^{'}}(\hat{x}) \frac{1}{\sqrt{2j^{'} + 1}} \braket{j m LM | j^{'} m^{'}} \frac{\delta\rho_{\alpha L, l^{'}j^{'}, lj}(r_{x}, r_{y}, \omega)}{r_{x} r_{y}} Y^{*}_{ljm}(\hat{y}), 
\end{align}
\begin{align}
\delta\rho_{\alpha LM}(\vecr,\omega) = Y_{LM}(\hat{\vecr})\delta\rho_{\alpha L}(r,\omega)/r^2.
\end{align}

Linear response equation, Eq.~(\ref{density_matrix_response}), for $\delta\rho_\alpha(\vecr,\omega)$ reduces to 
\begin{align} \label{density_response}
\delta\rho_{\alpha L}(r_x,\omega) &= \sum_{lj, l^{'}j^{'}} \Big\{ \frac{|\langle l^{'}j^{'} || Y_{L} || lj \rangle|^{2}}{2L + 1} \int dr_{x^{'}} \sum_{\beta} R^{\alpha \beta}_{0, l^{'}j^{'}, lj}(r_{x}, r_{x}; r_{x^{'}}, r_{x^{'}}; \omega) \sum_\gamma \frac{\del U_\beta[\rho]}{\del\rho_\gamma} \frac{1}{r^{2}_{x^{'}}} \delta {\rho}_{\gamma L}(r_{x^{'}}, \omega) \notag \\
&+ \frac{\langle l^{'}j^{'} || Y_{L} || lj \rangle}{2L + 1} \int dr_{x^{'}} dr_{y^{'}} \sum_{\beta} R^{\alpha \beta}_{0, l^{'}j^{'}, lj}(r_{x}, r_{x}; r_{y^{'}}, r_{x^{'}}; \omega) F_{\beta L, l^{'}j^{'}, lj}(r_{x^{'}}, r_{y^{'}}) \Big\}.
\end{align}

We evaluate the transition densities of a QRPA excited state $\ket{\nu}$ by using the linear response equation where
the operator $\hat{F}_{LM}$ in Eq.~(\ref{density_response}) is replaced with the multipole operator
$\hat{M}_{\lambda_{\nu} \mu_{\nu}}$.
\begin{align}
\rho^{({\rm tr}) \nu}_{\alpha \lambda_{\nu}}(r) = -\frac{C}{\pi r^{2}} {\rm Im} \delta \rho_{\alpha \lambda_{\nu}}(r,\omega_\nu),
\end{align}
where
\begin{align}
C = \frac{\sqrt{\int^{\hbar \omega_{\nu} + \Delta E}_{\hbar \omega_{\nu} - \Delta E} \hbar d\omega S(M_{\lambda_\nu}; \hbar \omega)}}{-\frac{1}{\pi}(2\lambda_{\nu}+1) \int dr \sum_{\alpha} f_{\alpha \lambda_{\nu}}(r_{x}) {\rm Im} \delta \rho_{\alpha \lambda_{\nu}}(r_{x}, \omega_\nu)}.
\end{align}
The integral interval $\Delta E$ is chosen to cover the peak corresponding to $\ket{\nu}$.
The same relation holds for the {\it pseudo transition density matrix},
\begin{align} \label{pseudo_transition_density_2}
\bar{\rho}^{({\rm tr}) \nu}_{\alpha \lambda_{\nu},l^{'}j^{'},lj}(r_x,r_{y}) = -\frac{C}{\pi r_{x} r_{y}} {\rm Im} \delta \bar{\rho}_{\alpha \lambda_{\nu},l^{'}j^{'},lj}(r_x,r_{y},\omega_\nu),
\end{align}
but with changing sign of the backward term in the unperturbed response function in Eq.~(\ref{density_response}) ($R_0 \to \bar{R}_0$, Ref.~\cite{Saito2021}).
We  compute the {\it pseudo transition density matrix} using Eq.~(\ref{pseudo_transition_density_2})
and the linear response equation with the Green's function instead of Eq.~(\ref{pseudo_transition_density}).

The matrix elements of ${\cal F}_{LM}(x,y)$ and the self-consistent field ${\cal V}^{\rm scf}_{LM}(x,y,\omega)$ are expanded as
\begin{align}
{\cal F}_{LM}(x, y) = \sum_{ljm, l^{'}j^{'}m^{'}} Y_{l^{'}j^{'}m^{'}}(\hat{x}) \frac{1}{\sqrt{2j^{'} + 1}} \braket{j m LM | j^{'} m^{'}} \frac{ {\cal F}_{L, l^{'}j^{'}, lj}(r_{x}, r_{y}, \omega)}{r_{x} r_{y}} Y^{*}_{ljm}(\hat{y}),
 \end{align} 
\begin{align}
{\cal V}^{\rm scf}_{LM}(x, y, \omega) = \sum_{ljm, l^{'}j^{'}m^{'}} Y_{l^{'}j^{'}m^{'}}(\hat{x}) \frac{1}{\sqrt{2j^{'} + 1}} \braket{j m LM | j^{'} m^{'}} \frac{{\cal V}^{\rm scf}_{L, l^{'}j^{'}, lj}(r_{x}, r_{y}, \omega)}{r_{x} r_{y}} Y^{*}_{ljm}(\hat{y}).
\end{align}

\bibliography{extended_DC_refs,extended_cQRPA_DC_refs}

\begin{thebibliography}{55}%
\makeatletter
\providecommand \@ifxundefined [1]{%
 \@ifx{#1\undefined}
}%
\providecommand \@ifnum [1]{%
 \ifnum #1\expandafter \@firstoftwo
 \else \expandafter \@secondoftwo
 \fi
}%
\providecommand \@ifx [1]{%
 \ifx #1\expandafter \@firstoftwo
 \else \expandafter \@secondoftwo
 \fi
}%
\providecommand \natexlab [1]{#1}%
\providecommand \enquote  [1]{``#1''}%
\providecommand \bibnamefont  [1]{#1}%
\providecommand \bibfnamefont [1]{#1}%
\providecommand \citenamefont [1]{#1}%
\providecommand \href@noop [0]{\@secondoftwo}%
\providecommand \href [0]{\begingroup \@sanitize@url \@href}%
\providecommand \@href[1]{\@@startlink{#1}\@@href}%
\providecommand \@@href[1]{\endgroup#1\@@endlink}%
\providecommand \@sanitize@url [0]{\catcode `\\12\catcode `\$12\catcode
  `\&12\catcode `\#12\catcode `\^12\catcode `\_12\catcode `\%12\relax}%
\providecommand \@@startlink[1]{}%
\providecommand \@@endlink[0]{}%
\providecommand \url  [0]{\begingroup\@sanitize@url \@url }%
\providecommand \@url [1]{\endgroup\@href {#1}{\urlprefix }}%
\providecommand \urlprefix  [0]{URL }%
\providecommand \Eprint [0]{\href }%
\providecommand \doibase [0]{http://dx.doi.org/}%
\providecommand \selectlanguage [0]{\@gobble}%
\providecommand \bibinfo  [0]{\@secondoftwo}%
\providecommand \bibfield  [0]{\@secondoftwo}%
\providecommand \translation [1]{[#1]}%
\providecommand \BibitemOpen [0]{}%
\providecommand \bibitemStop [0]{}%
\providecommand \bibitemNoStop [0]{.\EOS\space}%
\providecommand \EOS [0]{\spacefactor3000\relax}%
\providecommand \BibitemShut  [1]{\csname bibitem#1\endcsname}%
\let\auto@bib@innerbib\@empty
\bibitem [{\citenamefont {Burbidge}\ \emph {et~al.}(1957)\citenamefont
  {Burbidge}, \citenamefont {Burbidge}, \citenamefont {Fowler},\ and\
  \citenamefont {Hoyle}}]{B2FH}%
  \BibitemOpen
  \bibfield  {author} {\bibinfo {author} {\bibfnamefont {E.~M.}\ \bibnamefont
  {Burbidge}}, \bibinfo {author} {\bibfnamefont {G.~R.}\ \bibnamefont
  {Burbidge}}, \bibinfo {author} {\bibfnamefont {W.~A.}\ \bibnamefont
  {Fowler}}, \ and\ \bibinfo {author} {\bibfnamefont {F.}~\bibnamefont
  {Hoyle}},\ }\href@noop {} {\bibfield  {journal} {\bibinfo  {journal} {Rev.
  Mod. Phys.}\ }\textbf {\bibinfo {volume} {29}},\ \bibinfo {pages} {547}
  (\bibinfo {year} {1957})}\BibitemShut {NoStop}%
\bibitem [{\citenamefont {Arnould}\ \emph {et~al.}(2007)\citenamefont
  {Arnould}, \citenamefont {Goriely},\ and\ \citenamefont
  {Takahashi}}]{Arnould2007}%
  \BibitemOpen
  \bibfield  {author} {\bibinfo {author} {\bibfnamefont {M.}~\bibnamefont
  {Arnould}}, \bibinfo {author} {\bibfnamefont {S.}~\bibnamefont {Goriely}}, \
  and\ \bibinfo {author} {\bibfnamefont {K.}~\bibnamefont {Takahashi}},\
  }\href@noop {} {\bibfield  {journal} {\bibinfo  {journal} {Phys. Rep.}\
  }\textbf {\bibinfo {volume} {450}},\ \bibinfo {pages} {97} (\bibinfo {year}
  {2007})}\BibitemShut {NoStop}%
\bibitem [{\citenamefont {Cowan}\ \emph {et~al.}(2021)\citenamefont {Cowan},
  \citenamefont {Sneden}, \citenamefont {Lawler}, \citenamefont {Aprahamian},
  \citenamefont {Wiescher}, \citenamefont {Langanke}, \citenamefont
  {Martinez-Pinedo},\ and\ \citenamefont {Thielemann}}]{Cowan2021}%
  \BibitemOpen
  \bibfield  {author} {\bibinfo {author} {\bibfnamefont {J.~J.}\ \bibnamefont
  {Cowan}}, \bibinfo {author} {\bibfnamefont {C.}~\bibnamefont {Sneden}},
  \bibinfo {author} {\bibfnamefont {J.~E.}\ \bibnamefont {Lawler}}, \bibinfo
  {author} {\bibfnamefont {A.}~\bibnamefont {Aprahamian}}, \bibinfo {author}
  {\bibfnamefont {M.}~\bibnamefont {Wiescher}}, \bibinfo {author}
  {\bibfnamefont {K.}~\bibnamefont {Langanke}}, \bibinfo {author}
  {\bibfnamefont {G.}~\bibnamefont {Martinez-Pinedo}}, \ and\ \bibinfo {author}
  {\bibfnamefont {F.}~\bibnamefont {Thielemann}},\ }\href@noop {} {\bibfield
  {journal} {\bibinfo  {journal} {Rev. Mod. Phys.}\ }\textbf {\bibinfo {volume}
  {93}},\ \bibinfo {pages} {015002} (\bibinfo {year} {2021})}\BibitemShut
  {NoStop}%
\bibitem [{\citenamefont {Abbott}\ \emph
  {et~al.}(2017{\natexlab{a}})\citenamefont {Abbott} \emph
  {et~al.}}]{Abbott2017}%
  \BibitemOpen
  \bibfield  {author} {\bibinfo {author} {\bibfnamefont {B.~P.}\ \bibnamefont
  {Abbott}} \emph {et~al.},\ }\href@noop {} {\bibfield  {journal} {\bibinfo
  {journal} {Phys. Rev. Lett.}\ }\textbf {\bibinfo {volume} {119}},\ \bibinfo
  {pages} {161101} (\bibinfo {year} {2017}{\natexlab{a}})}\BibitemShut
  {NoStop}%
\bibitem [{\citenamefont {Abbott}\ \emph
  {et~al.}(2017{\natexlab{b}})\citenamefont {Abbott} \emph
  {et~al.}}]{Abbott2017_2}%
  \BibitemOpen
  \bibfield  {author} {\bibinfo {author} {\bibfnamefont {B.~P.}\ \bibnamefont
  {Abbott}} \emph {et~al.},\ }\href@noop {} {\bibfield  {journal} {\bibinfo
  {journal} {Astrophys. J. Lett.}\ }\textbf {\bibinfo {volume} {848}},\
  \bibinfo {pages} {L12} (\bibinfo {year} {2017}{\natexlab{b}})}\BibitemShut
  {NoStop}%
\bibitem [{\citenamefont {Pian}\ \emph {et~al.}(2017)\citenamefont {Pian} \emph
  {et~al.}}]{Pian2017}%
  \BibitemOpen
  \bibfield  {author} {\bibinfo {author} {\bibfnamefont {E.}~\bibnamefont
  {Pian}} \emph {et~al.},\ }\href@noop {} {\bibfield  {journal} {\bibinfo
  {journal} {nature}\ }\textbf {\bibinfo {volume} {551}},\ \bibinfo {pages}
  {67} (\bibinfo {year} {2017})}\BibitemShut {NoStop}%
\bibitem [{\citenamefont {Kasen}\ \emph {et~al.}(2017)\citenamefont {Kasen},
  \citenamefont {Metzger}, \citenamefont {Barnes}, \citenamefont {Quataert},\
  and\ \citenamefont {Ramirez-Ruiz}}]{Kasen2017}%
  \BibitemOpen
  \bibfield  {author} {\bibinfo {author} {\bibfnamefont {D.}~\bibnamefont
  {Kasen}}, \bibinfo {author} {\bibfnamefont {B.}~\bibnamefont {Metzger}},
  \bibinfo {author} {\bibfnamefont {J.}~\bibnamefont {Barnes}}, \bibinfo
  {author} {\bibfnamefont {E.}~\bibnamefont {Quataert}}, \ and\ \bibinfo
  {author} {\bibfnamefont {E.}~\bibnamefont {Ramirez-Ruiz}},\ }\href@noop {}
  {\bibfield  {journal} {\bibinfo  {journal} {Nature}\ }\textbf {\bibinfo
  {volume} {551}},\ \bibinfo {pages} {80} (\bibinfo {year} {2017})}\BibitemShut
  {NoStop}%
\bibitem [{\citenamefont {Villar}\ \emph {et~al.}(2017)\citenamefont {Villar}
  \emph {et~al.}}]{Villar2017}%
  \BibitemOpen
  \bibfield  {author} {\bibinfo {author} {\bibfnamefont {V.~A.}\ \bibnamefont
  {Villar}} \emph {et~al.},\ }\href@noop {} {\bibfield  {journal} {\bibinfo
  {journal} {Astrophys. J. Lett.}\ }\textbf {\bibinfo {volume} {851}},\
  \bibinfo {pages} {L21} (\bibinfo {year} {2017})}\BibitemShut {NoStop}%
\bibitem [{\citenamefont {Hauser}\ and\ \citenamefont
  {Feshbach}(1952)}]{Hauser1952}%
  \BibitemOpen
  \bibfield  {author} {\bibinfo {author} {\bibfnamefont {W.}~\bibnamefont
  {Hauser}}\ and\ \bibinfo {author} {\bibfnamefont {H.}~\bibnamefont
  {Feshbach}},\ }\href@noop {} {\bibfield  {journal} {\bibinfo  {journal}
  {Phys. Rev.}\ }\textbf {\bibinfo {volume} {87}},\ \bibinfo {pages} {366}
  (\bibinfo {year} {1952})}\BibitemShut {NoStop}%
\bibitem [{\citenamefont {k{\"a}ppeler}\ \emph {et~al.}(2011)\citenamefont
  {k{\"a}ppeler}, \citenamefont {Gallino}, \citenamefont {Bisterzo},\ and\
  \citenamefont {Aoki}}]{Kappeler2011}%
  \BibitemOpen
  \bibfield  {author} {\bibinfo {author} {\bibfnamefont {F.}~\bibnamefont
  {k{\"a}ppeler}}, \bibinfo {author} {\bibfnamefont {R.}~\bibnamefont
  {Gallino}}, \bibinfo {author} {\bibfnamefont {S.}~\bibnamefont {Bisterzo}}, \
  and\ \bibinfo {author} {\bibfnamefont {W.}~\bibnamefont {Aoki}},\ }\href@noop
  {} {\bibfield  {journal} {\bibinfo  {journal} {Rev. Mod. Phys.}\ }\textbf
  {\bibinfo {volume} {83}},\ \bibinfo {pages} {157} (\bibinfo {year}
  {2011})}\BibitemShut {NoStop}%
\bibitem [{\citenamefont {Mathews}\ \emph {et~al.}(1983)\citenamefont
  {Mathews}, \citenamefont {Mengoni}, \citenamefont {Thielemann},\ and\
  \citenamefont {Fowler}}]{Mathews1983}%
  \BibitemOpen
  \bibfield  {author} {\bibinfo {author} {\bibfnamefont {G.~J.}\ \bibnamefont
  {Mathews}}, \bibinfo {author} {\bibfnamefont {A.}~\bibnamefont {Mengoni}},
  \bibinfo {author} {\bibfnamefont {F.-K.}\ \bibnamefont {Thielemann}}, \ and\
  \bibinfo {author} {\bibfnamefont {W.~A.}\ \bibnamefont {Fowler}},\
  }\href@noop {} {\bibfield  {journal} {\bibinfo  {journal} {Astrophys. J.}\
  }\textbf {\bibinfo {volume} {270}},\ \bibinfo {pages} {740} (\bibinfo {year}
  {1983})}\BibitemShut {NoStop}%
\bibitem [{\citenamefont {Rauscher}\ \emph {et~al.}(1998)\citenamefont
  {Rauscher}, \citenamefont {Bieber}, \citenamefont {Oberhummer}, \citenamefont
  {Kratz}, \citenamefont {Dobaczewski}, \citenamefont {M{\"o}ller},\ and\
  \citenamefont {Sharma}}]{Rauscher1998}%
  \BibitemOpen
  \bibfield  {author} {\bibinfo {author} {\bibfnamefont {T.}~\bibnamefont
  {Rauscher}}, \bibinfo {author} {\bibfnamefont {R.}~\bibnamefont {Bieber}},
  \bibinfo {author} {\bibfnamefont {H.}~\bibnamefont {Oberhummer}}, \bibinfo
  {author} {\bibfnamefont {K.-L.}\ \bibnamefont {Kratz}}, \bibinfo {author}
  {\bibfnamefont {J.}~\bibnamefont {Dobaczewski}}, \bibinfo {author}
  {\bibfnamefont {P.}~\bibnamefont {M{\"o}ller}}, \ and\ \bibinfo {author}
  {\bibfnamefont {M.~M.}\ \bibnamefont {Sharma}},\ }\href@noop {} {\bibfield
  {journal} {\bibinfo  {journal} {Phys. Rev. C}\ }\textbf {\bibinfo {volume}
  {57}},\ \bibinfo {pages} {2031} (\bibinfo {year} {1998})}\BibitemShut
  {NoStop}%
\bibitem [{\citenamefont {Bonneau}\ \emph {et~al.}(2007)\citenamefont
  {Bonneau}, \citenamefont {Kawano}, \citenamefont {Watanabe},\ and\
  \citenamefont {Chiba}}]{Bonneau2007}%
  \BibitemOpen
  \bibfield  {author} {\bibinfo {author} {\bibfnamefont {L.}~\bibnamefont
  {Bonneau}}, \bibinfo {author} {\bibfnamefont {T.}~\bibnamefont {Kawano}},
  \bibinfo {author} {\bibfnamefont {T.}~\bibnamefont {Watanabe}}, \ and\
  \bibinfo {author} {\bibfnamefont {S.}~\bibnamefont {Chiba}},\ }\href@noop {}
  {\bibfield  {journal} {\bibinfo  {journal} {Phys. Rev. C}\ }\textbf {\bibinfo
  {volume} {75}},\ \bibinfo {pages} {054618} (\bibinfo {year}
  {2007})}\BibitemShut {NoStop}%
\bibitem [{\citenamefont {Chiba}\ \emph {et~al.}(2008)\citenamefont {Chiba},
  \citenamefont {Koura}, \citenamefont {Hayakawa}, \citenamefont {Maruyama},
  \citenamefont {Kawano},\ and\ \citenamefont {Kajino}}]{Chiba2008}%
  \BibitemOpen
  \bibfield  {author} {\bibinfo {author} {\bibfnamefont {S.}~\bibnamefont
  {Chiba}}, \bibinfo {author} {\bibfnamefont {H.}~\bibnamefont {Koura}},
  \bibinfo {author} {\bibfnamefont {T.}~\bibnamefont {Hayakawa}}, \bibinfo
  {author} {\bibfnamefont {T.}~\bibnamefont {Maruyama}}, \bibinfo {author}
  {\bibfnamefont {T.}~\bibnamefont {Kawano}}, \ and\ \bibinfo {author}
  {\bibfnamefont {T.}~\bibnamefont {Kajino}},\ }\href@noop {} {\bibfield
  {journal} {\bibinfo  {journal} {Phys. Rev. C}\ }\textbf {\bibinfo {volume}
  {77}},\ \bibinfo {pages} {015809} (\bibinfo {year} {2008})}\BibitemShut
  {NoStop}%
\bibitem [{\citenamefont {Rauscher}(2010)}]{Rauscher2010}%
  \BibitemOpen
  \bibfield  {author} {\bibinfo {author} {\bibfnamefont {T.}~\bibnamefont
  {Rauscher}},\ }\href@noop {} {\bibfield  {journal} {\bibinfo  {journal}
  {Nucl. Phys. A}\ }\textbf {\bibinfo {volume} {834}},\ \bibinfo {pages} {635c}
  (\bibinfo {year} {2010})}\BibitemShut {NoStop}%
\bibitem [{\citenamefont {Xu}\ and\ \citenamefont {Goriely}(2012)}]{Xu2012}%
  \BibitemOpen
  \bibfield  {author} {\bibinfo {author} {\bibfnamefont {Y.}~\bibnamefont
  {Xu}}\ and\ \bibinfo {author} {\bibfnamefont {S.}~\bibnamefont {Goriely}},\
  }\href@noop {} {\bibfield  {journal} {\bibinfo  {journal} {Phys. Rev. C}\
  }\textbf {\bibinfo {volume} {86}},\ \bibinfo {pages} {045801} (\bibinfo
  {year} {2012})}\BibitemShut {NoStop}%
\bibitem [{\citenamefont {Xu}\ \emph {et~al.}(2014)\citenamefont {Xu},
  \citenamefont {Goriely}, \citenamefont {Koning},\ and\ \citenamefont
  {Hilaire}}]{Xu2014}%
  \BibitemOpen
  \bibfield  {author} {\bibinfo {author} {\bibfnamefont {Y.}~\bibnamefont
  {Xu}}, \bibinfo {author} {\bibfnamefont {S.}~\bibnamefont {Goriely}},
  \bibinfo {author} {\bibfnamefont {A.~J.}\ \bibnamefont {Koning}}, \ and\
  \bibinfo {author} {\bibfnamefont {S.}~\bibnamefont {Hilaire}},\ }\href@noop
  {} {\bibfield  {journal} {\bibinfo  {journal} {Phys. Rev. C}\ }\textbf
  {\bibinfo {volume} {90}},\ \bibinfo {pages} {024604} (\bibinfo {year}
  {2014})}\BibitemShut {NoStop}%
\bibitem [{\citenamefont {Zhang}\ \emph {et~al.}(2015)\citenamefont {Zhang},
  \citenamefont {Peng}, \citenamefont {Smith}, \citenamefont {Arbanas},\ and\
  \citenamefont {Kozub}}]{Zhang2015}%
  \BibitemOpen
  \bibfield  {author} {\bibinfo {author} {\bibfnamefont {S.-S.}\ \bibnamefont
  {Zhang}}, \bibinfo {author} {\bibfnamefont {J.-P.}\ \bibnamefont {Peng}},
  \bibinfo {author} {\bibfnamefont {M.~S.}\ \bibnamefont {Smith}}, \bibinfo
  {author} {\bibfnamefont {G.}~\bibnamefont {Arbanas}}, \ and\ \bibinfo
  {author} {\bibfnamefont {R.~L.}\ \bibnamefont {Kozub}},\ }\href@noop {}
  {\bibfield  {journal} {\bibinfo  {journal} {Phys. Rev. C}\ }\textbf {\bibinfo
  {volume} {91}},\ \bibinfo {pages} {045802} (\bibinfo {year}
  {2015})}\BibitemShut {NoStop}%
\bibitem [{\citenamefont {Sieja}\ and\ \citenamefont
  {Goriely}(2021)}]{Sieja2021}%
  \BibitemOpen
  \bibfield  {author} {\bibinfo {author} {\bibfnamefont {K.}~\bibnamefont
  {Sieja}}\ and\ \bibinfo {author} {\bibfnamefont {S.}~\bibnamefont
  {Goriely}},\ }\href@noop {} {\bibfield  {journal} {\bibinfo  {journal} {Eur.
  Phys. J. A}\ }\textbf {\bibinfo {volume} {57}},\ \bibinfo {pages} {110}
  (\bibinfo {year} {2021})}\BibitemShut {NoStop}%
\bibitem [{\citenamefont {Tanihata}\ \emph {et~al.}(1985)\citenamefont
  {Tanihata} \emph {et~al.}}]{Tanihata1985}%
  \BibitemOpen
  \bibfield  {author} {\bibinfo {author} {\bibfnamefont {I.}~\bibnamefont
  {Tanihata}} \emph {et~al.},\ }\href@noop {} {\bibfield  {journal} {\bibinfo
  {journal} {Phys. Rev. Lett.}\ }\textbf {\bibinfo {volume} {55}},\ \bibinfo
  {pages} {2676} (\bibinfo {year} {1985})}\BibitemShut {NoStop}%
\bibitem [{\citenamefont {Hansen}\ and\ \citenamefont
  {Jonson}(1987)}]{Hansen1987}%
  \BibitemOpen
  \bibfield  {author} {\bibinfo {author} {\bibfnamefont {P.~G.}\ \bibnamefont
  {Hansen}}\ and\ \bibinfo {author} {\bibfnamefont {B.}~\bibnamefont
  {Jonson}},\ }\href@noop {} {\bibfield  {journal} {\bibinfo  {journal}
  {Europhys. Lett.}\ }\textbf {\bibinfo {volume} {4}},\ \bibinfo {pages} {409}
  (\bibinfo {year} {1987})}\BibitemShut {NoStop}%
\bibitem [{\citenamefont {Suzuki}\ \emph {et~al.}(1990)\citenamefont {Suzuki},
  \citenamefont {Ikeda},\ and\ \citenamefont {Sato}}]{Suzuki1990}%
  \BibitemOpen
  \bibfield  {author} {\bibinfo {author} {\bibfnamefont {Y.}~\bibnamefont
  {Suzuki}}, \bibinfo {author} {\bibfnamefont {K.}~\bibnamefont {Ikeda}}, \
  and\ \bibinfo {author} {\bibfnamefont {H.}~\bibnamefont {Sato}},\ }\href@noop
  {} {\bibfield  {journal} {\bibinfo  {journal} {Prog. Theor. Phys.}\ }\textbf
  {\bibinfo {volume} {83}},\ \bibinfo {pages} {180} (\bibinfo {year}
  {1990})}\BibitemShut {NoStop}%
\bibitem [{\citenamefont {Bertsch}\ and\ \citenamefont
  {Esbensen}(1991)}]{Bertsch1991}%
  \BibitemOpen
  \bibfield  {author} {\bibinfo {author} {\bibfnamefont {G.~F.}\ \bibnamefont
  {Bertsch}}\ and\ \bibinfo {author} {\bibfnamefont {H.}~\bibnamefont
  {Esbensen}},\ }\href@noop {} {\bibfield  {journal} {\bibinfo  {journal} {Ann.
  Phys. A}\ }\textbf {\bibinfo {volume} {209}},\ \bibinfo {pages} {327}
  (\bibinfo {year} {1991})}\BibitemShut {NoStop}%
\bibitem [{\citenamefont {Paar}\ \emph {et~al.}(2007)\citenamefont {Paar},
  \citenamefont {Vretenar}, \citenamefont {Khan},\ and\ \citenamefont
  {Col{\`o}}}]{Paar2007}%
  \BibitemOpen
  \bibfield  {author} {\bibinfo {author} {\bibfnamefont {N.}~\bibnamefont
  {Paar}}, \bibinfo {author} {\bibfnamefont {D.}~\bibnamefont {Vretenar}},
  \bibinfo {author} {\bibfnamefont {E.}~\bibnamefont {Khan}}, \ and\ \bibinfo
  {author} {\bibfnamefont {G.}~\bibnamefont {Col{\`o}}},\ }\href@noop {}
  {\bibfield  {journal} {\bibinfo  {journal} {Rep. Prog. Phys.}\ }\textbf
  {\bibinfo {volume} {70}},\ \bibinfo {pages} {691} (\bibinfo {year}
  {2007})}\BibitemShut {NoStop}%
\bibitem [{\citenamefont {Tanihata}\ \emph {et~al.}(2013)\citenamefont
  {Tanihata}, \citenamefont {Savajols},\ and\ \citenamefont
  {Kanungo}}]{Tanihata2013}%
  \BibitemOpen
  \bibfield  {author} {\bibinfo {author} {\bibfnamefont {I.}~\bibnamefont
  {Tanihata}}, \bibinfo {author} {\bibfnamefont {H.}~\bibnamefont {Savajols}},
  \ and\ \bibinfo {author} {\bibfnamefont {R.}~\bibnamefont {Kanungo}},\
  }\href@noop {} {\bibfield  {journal} {\bibinfo  {journal} {Prog. Part. Nucl.
  Phys.}\ }\textbf {\bibinfo {volume} {68}},\ \bibinfo {pages} {215} (\bibinfo
  {year} {2013})}\BibitemShut {NoStop}%
\bibitem [{\citenamefont {Goriely}(1998)}]{Goriely1998}%
  \BibitemOpen
  \bibfield  {author} {\bibinfo {author} {\bibfnamefont {S.}~\bibnamefont
  {Goriely}},\ }\href@noop {} {\bibfield  {journal} {\bibinfo  {journal} {Phys.
  Lett. B}\ }\textbf {\bibinfo {volume} {436}},\ \bibinfo {pages} {10}
  (\bibinfo {year} {1998})}\BibitemShut {NoStop}%
\bibitem [{\citenamefont {Goriely}\ and\ \citenamefont
  {Khan}(2002)}]{Goriely2002}%
  \BibitemOpen
  \bibfield  {author} {\bibinfo {author} {\bibfnamefont {S.}~\bibnamefont
  {Goriely}}\ and\ \bibinfo {author} {\bibfnamefont {E.}~\bibnamefont {Khan}},\
  }\href@noop {} {\bibfield  {journal} {\bibinfo  {journal} {Nucl. Phys.}\
  }\textbf {\bibinfo {volume} {A706}},\ \bibinfo {pages} {217} (\bibinfo {year}
  {2002})}\BibitemShut {NoStop}%
\bibitem [{\citenamefont {Goriely}\ \emph {et~al.}(2004)\citenamefont
  {Goriely}, \citenamefont {Khan},\ and\ \citenamefont {Shamyn}}]{Goriely2004}%
  \BibitemOpen
  \bibfield  {author} {\bibinfo {author} {\bibfnamefont {S.}~\bibnamefont
  {Goriely}}, \bibinfo {author} {\bibfnamefont {E.}~\bibnamefont {Khan}}, \
  and\ \bibinfo {author} {\bibfnamefont {M.}~\bibnamefont {Shamyn}},\
  }\href@noop {} {\bibfield  {journal} {\bibinfo  {journal} {Nucl. Phys. A}\
  }\textbf {\bibinfo {volume} {739}},\ \bibinfo {pages} {331} (\bibinfo {year}
  {2004})}\BibitemShut {NoStop}%
\bibitem [{\citenamefont {Litvinova}\ \emph {et~al.}(2009)\citenamefont
  {Litvinova}, \citenamefont {Loens}, \citenamefont {Langanke}, \citenamefont
  {Mart\'{i}nez-Pinedo}, \citenamefont {Rauscher}, \citenamefont {Ring},
  \citenamefont {Thielemann},\ and\ \citenamefont {Tselyaev}}]{Litvinova2009}%
  \BibitemOpen
  \bibfield  {author} {\bibinfo {author} {\bibfnamefont {E.}~\bibnamefont
  {Litvinova}}, \bibinfo {author} {\bibfnamefont {H.~P.}\ \bibnamefont
  {Loens}}, \bibinfo {author} {\bibfnamefont {K.}~\bibnamefont {Langanke}},
  \bibinfo {author} {\bibfnamefont {G.}~\bibnamefont {Mart\'{i}nez-Pinedo}},
  \bibinfo {author} {\bibfnamefont {T.}~\bibnamefont {Rauscher}}, \bibinfo
  {author} {\bibfnamefont {P.}~\bibnamefont {Ring}}, \bibinfo {author}
  {\bibfnamefont {F.-K.}\ \bibnamefont {Thielemann}}, \ and\ \bibinfo {author}
  {\bibfnamefont {V.}~\bibnamefont {Tselyaev}},\ }\href@noop {} {\bibfield
  {journal} {\bibinfo  {journal} {Nucl. Phys. A}\ }\textbf {\bibinfo {volume}
  {823}},\ \bibinfo {pages} {26} (\bibinfo {year} {2009})}\BibitemShut
  {NoStop}%
\bibitem [{\citenamefont {Avdeenkov}\ \emph {et~al.}(2011)\citenamefont
  {Avdeenkov}, \citenamefont {Goriely}, \citenamefont {Kamerdzhiev},\ and\
  \citenamefont {Krewald}}]{Avdeenkov2011}%
  \BibitemOpen
  \bibfield  {author} {\bibinfo {author} {\bibfnamefont {A.}~\bibnamefont
  {Avdeenkov}}, \bibinfo {author} {\bibfnamefont {S.}~\bibnamefont {Goriely}},
  \bibinfo {author} {\bibfnamefont {S.}~\bibnamefont {Kamerdzhiev}}, \ and\
  \bibinfo {author} {\bibfnamefont {S.}~\bibnamefont {Krewald}},\ }\href@noop
  {} {\bibfield  {journal} {\bibinfo  {journal} {Phys. Rev. C}\ }\textbf
  {\bibinfo {volume} {83}},\ \bibinfo {pages} {064316} (\bibinfo {year}
  {2011})}\BibitemShut {NoStop}%
\bibitem [{\citenamefont {Daoutidis}\ and\ \citenamefont
  {Goriely}(2012)}]{Daoutidis2012}%
  \BibitemOpen
  \bibfield  {author} {\bibinfo {author} {\bibfnamefont {I.}~\bibnamefont
  {Daoutidis}}\ and\ \bibinfo {author} {\bibfnamefont {S.}~\bibnamefont
  {Goriely}},\ }\href@noop {} {\bibfield  {journal} {\bibinfo  {journal} {Phys.
  Rev. C}\ }\textbf {\bibinfo {volume} {86}},\ \bibinfo {pages} {034328}
  (\bibinfo {year} {2012})}\BibitemShut {NoStop}%
\bibitem [{\citenamefont {Tsoneva}\ \emph {et~al.}(2015)\citenamefont
  {Tsoneva}, \citenamefont {Goriely}, \citenamefont {Lenske},\ and\
  \citenamefont {Schwengner}}]{Tsoneva2015}%
  \BibitemOpen
  \bibfield  {author} {\bibinfo {author} {\bibfnamefont {N.}~\bibnamefont
  {Tsoneva}}, \bibinfo {author} {\bibfnamefont {S.}~\bibnamefont {Goriely}},
  \bibinfo {author} {\bibfnamefont {H.}~\bibnamefont {Lenske}}, \ and\ \bibinfo
  {author} {\bibfnamefont {R.}~\bibnamefont {Schwengner}},\ }\href@noop {}
  {\bibfield  {journal} {\bibinfo  {journal} {Phys. Rev. C}\ }\textbf {\bibinfo
  {volume} {91}},\ \bibinfo {pages} {044318} (\bibinfo {year}
  {2015})}\BibitemShut {NoStop}%
\bibitem [{\citenamefont {Martini}\ \emph {et~al.}(2016)\citenamefont
  {Martini}, \citenamefont {P\'{e}ru}, \citenamefont {Hilaire}, \citenamefont
  {Goriely},\ and\ \citenamefont {Lechaftois}}]{Martini2016}%
  \BibitemOpen
  \bibfield  {author} {\bibinfo {author} {\bibfnamefont {M.}~\bibnamefont
  {Martini}}, \bibinfo {author} {\bibfnamefont {S.}~\bibnamefont {P\'{e}ru}},
  \bibinfo {author} {\bibfnamefont {S.}~\bibnamefont {Hilaire}}, \bibinfo
  {author} {\bibfnamefont {S.}~\bibnamefont {Goriely}}, \ and\ \bibinfo
  {author} {\bibfnamefont {F.}~\bibnamefont {Lechaftois}},\ }\href@noop {}
  {\bibfield  {journal} {\bibinfo  {journal} {Phys. Rev. C}\ }\textbf {\bibinfo
  {volume} {94}},\ \bibinfo {pages} {014304} (\bibinfo {year}
  {2016})}\BibitemShut {NoStop}%
\bibitem [{\citenamefont {Matsuo}(2015)}]{Matsuo2015}%
  \BibitemOpen
  \bibfield  {author} {\bibinfo {author} {\bibfnamefont {M.}~\bibnamefont
  {Matsuo}},\ }\href@noop {} {\bibfield  {journal} {\bibinfo  {journal} {Phys.
  Rev. C}\ }\textbf {\bibinfo {volume} {91}},\ \bibinfo {pages} {034604}
  (\bibinfo {year} {2015})}\BibitemShut {NoStop}%
\bibitem [{\citenamefont {Saito}\ and\ \citenamefont
  {Matsuo}(2023)}]{Saito2023}%
  \BibitemOpen
  \bibfield  {author} {\bibinfo {author} {\bibfnamefont {T.}~\bibnamefont
  {Saito}}\ and\ \bibinfo {author} {\bibfnamefont {M.}~\bibnamefont {Matsuo}},\
  }\href@noop {} {\bibfield  {journal} {\bibinfo  {journal} {Phys. Rev. C}\
  }\textbf {\bibinfo {volume} {107}},\ \bibinfo {pages} {064607} (\bibinfo
  {year} {2023})}\BibitemShut {NoStop}%
\bibitem [{\citenamefont {Zangwill}\ and\ \citenamefont
  {Soven}(1980)}]{Zangwill1980}%
  \BibitemOpen
  \bibfield  {author} {\bibinfo {author} {\bibfnamefont {A.}~\bibnamefont
  {Zangwill}}\ and\ \bibinfo {author} {\bibfnamefont {P.}~\bibnamefont
  {Soven}},\ }\href@noop {} {\bibfield  {journal} {\bibinfo  {journal} {Phys.
  Rev. A}\ }\textbf {\bibinfo {volume} {21}},\ \bibinfo {pages} {1561}
  (\bibinfo {year} {1980})}\BibitemShut {NoStop}%
\bibitem [{\citenamefont {Ring}\ and\ \citenamefont {Schuck}(1980)}]{Ring1980}%
  \BibitemOpen
  \bibfield  {author} {\bibinfo {author} {\bibfnamefont {P.}~\bibnamefont
  {Ring}}\ and\ \bibinfo {author} {\bibfnamefont {P.}~\bibnamefont {Schuck}},\
  }\href@noop {} {\emph {\bibinfo {title} {The Nuclear Many-Body Problem}}}\
  (\bibinfo  {publisher} {Springer-Verlag, Berlin},\ \bibinfo {year}
  {1980})\BibitemShut {NoStop}%
\bibitem [{\citenamefont {Bertulani}\ and\ \citenamefont
  {Danielewicz}(2004)}]{Bertulani2004}%
  \BibitemOpen
  \bibfield  {author} {\bibinfo {author} {\bibfnamefont {C.~A.}\ \bibnamefont
  {Bertulani}}\ and\ \bibinfo {author} {\bibfnamefont {P.}~\bibnamefont
  {Danielewicz}},\ }\href@noop {} {\emph {\bibinfo {title} {Introduction to
  Nuclear Reaction}}}\ (\bibinfo  {publisher} {Institute of Physics Publishing,
  London},\ \bibinfo {year} {2004})\BibitemShut {NoStop}%
\bibitem [{\citenamefont {Thompson}\ and\ \citenamefont
  {Nunes}(2004)}]{Thompson2009}%
  \BibitemOpen
  \bibfield  {author} {\bibinfo {author} {\bibfnamefont {I.~J.}\ \bibnamefont
  {Thompson}}\ and\ \bibinfo {author} {\bibfnamefont {F.~M.}\ \bibnamefont
  {Nunes}},\ }\href@noop {} {\emph {\bibinfo {title} {Nuclear Reactions for
  Astrophysics}}}\ (\bibinfo  {publisher} {Cambridge University Press,
  Cambridge},\ \bibinfo {year} {2004})\BibitemShut {NoStop}%
\bibitem [{\citenamefont {Saito}\ and\ \citenamefont
  {Matsuo}(2021)}]{Saito2021}%
  \BibitemOpen
  \bibfield  {author} {\bibinfo {author} {\bibfnamefont {T.}~\bibnamefont
  {Saito}}\ and\ \bibinfo {author} {\bibfnamefont {M.}~\bibnamefont {Matsuo}},\
  }\href@noop {} {\bibfield  {journal} {\bibinfo  {journal} {Phys. Rev. C}\
  }\textbf {\bibinfo {volume} {104}},\ \bibinfo {pages} {034305} (\bibinfo
  {year} {2021})}\BibitemShut {NoStop}%
\bibitem [{\citenamefont {Matsuo}(2001)}]{Matsuo2001}%
  \BibitemOpen
  \bibfield  {author} {\bibinfo {author} {\bibfnamefont {M.}~\bibnamefont
  {Matsuo}},\ }\href@noop {} {\bibfield  {journal} {\bibinfo  {journal} {Nucl.
  Phys. A}\ }\textbf {\bibinfo {volume} {696}},\ \bibinfo {pages} {371}
  (\bibinfo {year} {2001})}\BibitemShut {NoStop}%
\bibitem [{\citenamefont {Shimoyama}\ and\ \citenamefont
  {Matsuo}(2013)}]{Shimoyama2013}%
  \BibitemOpen
  \bibfield  {author} {\bibinfo {author} {\bibfnamefont {H.}~\bibnamefont
  {Shimoyama}}\ and\ \bibinfo {author} {\bibfnamefont {M.}~\bibnamefont
  {Matsuo}},\ }\href@noop {} {\bibfield  {journal} {\bibinfo  {journal} {Phys.
  Rev. C}\ }\textbf {\bibinfo {volume} {88}},\ \bibinfo {pages} {054308}
  (\bibinfo {year} {2013})}\BibitemShut {NoStop}%
\bibitem [{\citenamefont {Dobaczewski}\ \emph {et~al.}(1984)\citenamefont
  {Dobaczewski}, \citenamefont {Flocard},\ and\ \citenamefont
  {Treiner}}]{Dobaczewski1984}%
  \BibitemOpen
  \bibfield  {author} {\bibinfo {author} {\bibfnamefont {J.}~\bibnamefont
  {Dobaczewski}}, \bibinfo {author} {\bibfnamefont {H.}~\bibnamefont
  {Flocard}}, \ and\ \bibinfo {author} {\bibfnamefont {J.}~\bibnamefont
  {Treiner}},\ }\href@noop {} {\bibfield  {journal} {\bibinfo  {journal} {Nucl.
  Phys.}\ }\textbf {\bibinfo {volume} {A422}},\ \bibinfo {pages} {103}
  (\bibinfo {year} {1984})}\BibitemShut {NoStop}%
\bibitem [{\citenamefont {Surman}\ \emph {et~al.}(2014)\citenamefont {Surman},
  \citenamefont {Mumpower}, \citenamefont {Sinclair}, \citenamefont {Jones},
  \citenamefont {Hix},\ and\ \citenamefont {McLaughlin}}]{Surman2014}%
  \BibitemOpen
  \bibfield  {author} {\bibinfo {author} {\bibfnamefont {R.}~\bibnamefont
  {Surman}}, \bibinfo {author} {\bibfnamefont {M.}~\bibnamefont {Mumpower}},
  \bibinfo {author} {\bibfnamefont {R.}~\bibnamefont {Sinclair}}, \bibinfo
  {author} {\bibfnamefont {L.~L.}\ \bibnamefont {Jones}}, \bibinfo {author}
  {\bibfnamefont {W.~R.}\ \bibnamefont {Hix}}, \ and\ \bibinfo {author}
  {\bibfnamefont {G.~C.}\ \bibnamefont {McLaughlin}},\ }\href@noop {}
  {\bibfield  {journal} {\bibinfo  {journal} {AIP Advances}\ }\textbf {\bibinfo
  {volume} {4}},\ \bibinfo {pages} {041008} (\bibinfo {year}
  {2014})}\BibitemShut {NoStop}%
\bibitem [{Mas()}]{Massexpl}%
  \BibitemOpen
  \href@noop {} {}\bibinfo {note} {Mass explorer,
  http://massexplorer.frib.msu.edu/content/DFTMassTables.html}\BibitemShut
  {NoStop}%
\bibitem [{\citenamefont {Bartel}\ and\ \citenamefont
  {Quentin}(1982)}]{Bartel1982}%
  \BibitemOpen
  \bibfield  {author} {\bibinfo {author} {\bibfnamefont {J.}~\bibnamefont
  {Bartel}}\ and\ \bibinfo {author} {\bibfnamefont {P.}~\bibnamefont
  {Quentin}},\ }\href@noop {} {\bibfield  {journal} {\bibinfo  {journal} {Nucl.
  Phys. A}\ }\textbf {\bibinfo {volume} {386}},\ \bibinfo {pages} {79}
  (\bibinfo {year} {1982})}\BibitemShut {NoStop}%
\bibitem [{\citenamefont {Chabanat}\ \emph {et~al.}(1998)\citenamefont
  {Chabanat}, \citenamefont {Bonche}, \citenamefont {Haensel}, \citenamefont
  {Meyer},\ and\ \citenamefont {Schaeffer}}]{Shabanat1998}%
  \BibitemOpen
  \bibfield  {author} {\bibinfo {author} {\bibfnamefont {E.}~\bibnamefont
  {Chabanat}}, \bibinfo {author} {\bibfnamefont {P.}~\bibnamefont {Bonche}},
  \bibinfo {author} {\bibfnamefont {P.}~\bibnamefont {Haensel}}, \bibinfo
  {author} {\bibfnamefont {J.}~\bibnamefont {Meyer}}, \ and\ \bibinfo {author}
  {\bibfnamefont {R.}~\bibnamefont {Schaeffer}},\ }\href@noop {} {\bibfield
  {journal} {\bibinfo  {journal} {Nucl. Phys. A}\ }\textbf {\bibinfo {volume}
  {635}},\ \bibinfo {pages} {231} (\bibinfo {year} {1998})}\BibitemShut
  {NoStop}%
\bibitem [{\citenamefont {Matsuo}\ and\ \citenamefont
  {Serizawa}(2010)}]{Matsuo2010}%
  \BibitemOpen
  \bibfield  {author} {\bibinfo {author} {\bibfnamefont {M.}~\bibnamefont
  {Matsuo}}\ and\ \bibinfo {author} {\bibfnamefont {Y.}~\bibnamefont
  {Serizawa}},\ }\href@noop {} {\bibfield  {journal} {\bibinfo  {journal}
  {Phys. Rev. C}\ }\textbf {\bibinfo {volume} {82}},\ \bibinfo {pages} {024318}
  (\bibinfo {year} {2010})}\BibitemShut {NoStop}%
\bibitem [{\citenamefont {Shand}\ \emph {et~al.}(2017)\citenamefont {Shand}
  \emph {et~al.}}]{Shand2017}%
  \BibitemOpen
  \bibfield  {author} {\bibinfo {author} {\bibfnamefont {C.}~\bibnamefont
  {Shand}} \emph {et~al.},\ }\href@noop {} {\bibfield  {journal} {\bibinfo
  {journal} {Phys. Lett. B}\ }\textbf {\bibinfo {volume} {773}},\ \bibinfo
  {pages} {492} (\bibinfo {year} {2017})}\BibitemShut {NoStop}%
\bibitem [{\citenamefont {Miernik}\ \emph {et~al.}(2013)\citenamefont {Miernik}
  \emph {et~al.}}]{Miernik2013}%
  \BibitemOpen
  \bibfield  {author} {\bibinfo {author} {\bibfnamefont {K.}~\bibnamefont
  {Miernik}} \emph {et~al.},\ }\href@noop {} {\bibfield  {journal} {\bibinfo
  {journal} {Phys. Rev. Lett.}\ }\textbf {\bibinfo {volume} {111}},\ \bibinfo
  {pages} {132502} (\bibinfo {year} {2013})}\BibitemShut {NoStop}%
\bibitem [{\citenamefont {Tsoneva}\ and\ \citenamefont
  {Lenske}(2011)}]{Tsoneva2011}%
  \BibitemOpen
  \bibfield  {author} {\bibinfo {author} {\bibfnamefont {N.}~\bibnamefont
  {Tsoneva}}\ and\ \bibinfo {author} {\bibfnamefont {H.}~\bibnamefont
  {Lenske}},\ }\href@noop {} {\bibfield  {journal} {\bibinfo  {journal} {Phys.
  Lett. B}\ }\textbf {\bibinfo {volume} {695}},\ \bibinfo {pages} {174}
  (\bibinfo {year} {2011})}\BibitemShut {NoStop}%
\bibitem [{\citenamefont {Matsuo}\ \emph {et~al.}(2005)\citenamefont {Matsuo},
  \citenamefont {Mizuyama},\ and\ \citenamefont {Serizawa}}]{Matsuo2005}%
  \BibitemOpen
  \bibfield  {author} {\bibinfo {author} {\bibfnamefont {M.}~\bibnamefont
  {Matsuo}}, \bibinfo {author} {\bibfnamefont {K.}~\bibnamefont {Mizuyama}}, \
  and\ \bibinfo {author} {\bibfnamefont {Y.}~\bibnamefont {Serizawa}},\
  }\href@noop {} {\bibfield  {journal} {\bibinfo  {journal} {Phys. Rev. C}\
  }\textbf {\bibinfo {volume} {71}},\ \bibinfo {pages} {064326} (\bibinfo
  {year} {2005})}\BibitemShut {NoStop}%
\bibitem [{\citenamefont {Dobaczewski}\ \emph {et~al.}(1996)\citenamefont
  {Dobaczewski}, \citenamefont {Nazarewicz}, \citenamefont {Werner},
  \citenamefont {Berger}, \citenamefont {Chinn},\ and\ \citenamefont
  {Decharge}}]{Dobaczewski1996}%
  \BibitemOpen
  \bibfield  {author} {\bibinfo {author} {\bibfnamefont {J.}~\bibnamefont
  {Dobaczewski}}, \bibinfo {author} {\bibfnamefont {W.}~\bibnamefont
  {Nazarewicz}}, \bibinfo {author} {\bibfnamefont {T.~R.}\ \bibnamefont
  {Werner}}, \bibinfo {author} {\bibfnamefont {J.~F.}\ \bibnamefont {Berger}},
  \bibinfo {author} {\bibfnamefont {C.~R.}\ \bibnamefont {Chinn}}, \ and\
  \bibinfo {author} {\bibfnamefont {J.}~\bibnamefont {Decharge}},\ }\href@noop
  {} {\bibfield  {journal} {\bibinfo  {journal} {Phys. Rev. C}\ }\textbf
  {\bibinfo {volume} {53}},\ \bibinfo {pages} {2809} (\bibinfo {year}
  {1996})}\BibitemShut {NoStop}%
\bibitem [{\citenamefont {Bennaceur}\ \emph {et~al.}(1999)\citenamefont
  {Bennaceur}, \citenamefont {Dobaczewski},\ and\ \citenamefont
  {Ploszajczak}}]{Bennaceur1999}%
  \BibitemOpen
  \bibfield  {author} {\bibinfo {author} {\bibfnamefont {K.}~\bibnamefont
  {Bennaceur}}, \bibinfo {author} {\bibfnamefont {J.}~\bibnamefont
  {Dobaczewski}}, \ and\ \bibinfo {author} {\bibfnamefont {M.}~\bibnamefont
  {Ploszajczak}},\ }\href@noop {} {\bibfield  {journal} {\bibinfo  {journal}
  {Phys. Rev. C}\ }\textbf {\bibinfo {volume} {60}},\ \bibinfo {pages} {034308}
  (\bibinfo {year} {1999})}\BibitemShut {NoStop}%
\bibitem [{\citenamefont {Balgac}()}]{Balgac2000}%
  \BibitemOpen
  \bibfield  {author} {\bibinfo {author} {\bibfnamefont {A.}~\bibnamefont
  {Balgac}},\ }\href@noop {} {\bibinfo  {journal} {arXiv:nucl-th/9907088}\
  }\BibitemShut {NoStop}%
\end{thebibliography}%

\end{document}